\documentclass{jfm}
\usepackage{graphicx,mathtools,url,lineno}

\newcommand{\infd}{\text{d}}

\newcommand{\pd}[2]{\frac{\partial{#1}}{\partial{#2}}}

\newcommand{\lagd}[3]{\frac{\text{D}_{#1}{#2}}{\text{D}{#3}}}
\newcommand{\Grad}{\boldsymbol{\nabla}}
\newcommand{\Div}{\Grad\cdot}

\newcommand{\delsq}{\nabla^2}
\newcommand{\trace}[1]{\text{tr}\left(#1\right)}
\newcommand{\omp}{(1-\phi)}
\newcommand{\vel}{\boldsymbol{v}}
\newcommand{\cmp}{\mathcal{C}}
\newcommand{\ndcl}{\mathcal{R}}
\newcommand{\strr}{\dot{\varepsilon}_{II}}
\newcommand{\strrten}{\dot{\boldsymbol{\varepsilon}}}
\newcommand{\identity}{\boldsymbol{I}}
\newcommand{\cmpvisc}{\zeta_\phi}
\newcommand{\cmpviscnl}{\widetilde{\zeta}_\phi}
\newcommand{\shearvisc}{\eta_\phi}
\newcommand{\shearviscnl}{\widetilde{\eta}_\phi}
\newcommand{\dilvisc}{D_\phi}
\newcommand{\dilviscnl}{\widetilde{D}_\phi}
\newcommand{\mobility}{M_\phi}
\newcommand{\torque}{\mathcal{T}}
\newcommand{\xhat}{\hat{\boldsymbol{x}}}
\newcommand{\yhat}{\hat{\boldsymbol{y}}}
\newcommand{\zhat}{\hat{\boldsymbol{z}}}
\newcommand{\rhat}{\hat{\boldsymbol{r}}}
\newcommand{\phihat}{\hat{\boldsymbol{\varphi}}}

\shorttitle{Dilatancy of partially molten rock}
\shortauthor{R.F. Katz, J.F. Rudge \& L.N. Hansen}

\title{Granular dilatancy and non-local fluidity of partially molten rock}

\author{Richard F.~Katz\aff{1}\corresp{\email{richard.katz@earth.ox.ac.uk}}, John F.~Rudge\aff{2} \and Lars N.~Hansen\aff{3}}

\affiliation{\aff{1}Department of Earth Sciences, University of Oxford, Oxford OX1 3AN, UK
\aff{2}Department of Earth Sciences, University of Cambridge, Cambridge CB3 0EZ, UK
\aff{2}Department of Earth \& Environmental Sciences, University of
Minnesota, Minneapolis 55455, USA}

\begin{document}

\maketitle

\begin{abstract}
Partially molten rock is a densely packed, melt-saturated, granular medium, but it has seldom been considered in these terms.  In this paper, we extend the continuum theory of partially molten rock to incorporate the physics of granular media. Our formulation includes dilatancy in a viscous constitutive law and introduces a non-local fluidity. We analyse the resulting poro-viscous--granular theory in terms of two modes of liquid--solid segregation that are observed in published torsion experiments: localisation of liquid into high-porosity sheets and radially inward liquid flow. We show that the newly incorporated granular physics brings the theory into agreement with experiments. We discuss these results in the context of grain-scale physics across the nominal jamming fraction at the high homologous temperatures relevant in geological systems. 
\end{abstract}

\begin{keywords}
Authors should not enter keywords on the manuscript, as these must be chosen by the author during the online submission process and will then be added during the typesetting process (see http://journals.cambridge.org/data/\linebreak[3]relatedlink/jfm-\linebreak[3]keywords.pdf for the full list)
\end{keywords}

\section{Introduction}

Partially molten rock is a physical system that is central to many geological and planetary processes.  It is a densely packed, melt-saturated, granular medium, but it has seldom been considered in these terms.  Continuum models of partially molten rock treat the solid and liquid phases as interpenetrating fluids in a poro-viscous, zero-Reynolds-number theory \citep[e.g.,][]{mckenzie1984generation, fowler1990compaction}. In such models, effects arising from the discrete grains are neglected, except insofar as they affect the creep viscosity.  For example, during Coble creep, the melt phase provides a fast pathway for mass diffusion around grains \citep[e.g.,][]{takei2009viscous, rudge2018viscosities}.  Deformation experiments on partially molten rock are typically parameterised in terms of an isotropic flow law with a weakening factor that depends on the volume fraction of melt within pores \citep[e.g.,][]{kohlstedt1996rheology, kelemen1997review}.  These physics were reviewed by \cite{katz2022physics}.

However, deformation of partially molten rock inevitably includes a component of sliding along grain boundaries \citep[e.g.,][]{hansen2011grain, rudge2021micropolar}. The granular origins and importance of such sliding in partially molten rock were recognised by \cite{paterson1995theory} and elaborated by \cite{paterson2001granular}. In those works, the geometric grain-compatibility problem arising from grain-boundary sliding is assumed to be entirely resolved by shape-change of the grains, which occurs by diffusion or lattice dislocations \citep{langdon2006grain}. A third possibility is noted by \cite{paterson2001granular} but then neglected: that incompatibility is resolved by relative motion of undeforming grains in a granular flow.  This mechanism is fundamental in the physics of athermal granular media \citep[e.g.,][]{forterre2008flows}; it gives rise to dilatancy and non-local granular fluidity.

In this paper, we extend the continuum theory of partially molten rock to incorporate the physics of a granular medium. As a hypothesis for the essential granular physics, we adapt and include theory for dilatancy and non-local fluidity.  We test this hypothesis by modelling published laboratory experiments in which partially molten rock is subjected to torsional deformation. The deformation drives liquid--solid segregation and yields robust patterns of melt localisation. We show that the inclusion of granular physics brings model predictions into agreement with laboratory data.

The laboratory experiments, detailed in \cite{king2010stress} and reviewed in \S\ref{sec:experiments} below, are conducted on synthetic rocks comprising solid olivine grains and liquid basaltic melt. Hot-pressed, nominally uniform, cylindrical samples of this aggregate are sheared in a torsion apparatus at high temperature and confining pressure.  During shear, two modes of liquid--solid segregation occur simultaneously.  The first is a pattern-forming localisation of the liquid into high-porosity sheets that form at 15--20$^\circ$ to the shear plane \citep{holtzman2003stress}. The sheets are typically measured in their cross section, where they appear as bands with a characteristic spacing.  The second is a radially inward porous flow of liquid, accommodated by a radially outward flow of solid \citep{qi2015experimental}.  

A satisfactory physical understanding of these flow phenomena has been elusive, though much has been learned through theoretical analysis. Localisation of the liquid phase into sheets at 45$^\circ$ to the shear plane was predicted by \cite{stevenson1989spontaneous} and \cite{spiegelman2003linear} to be a consequence of a porosity-weakening viscosity of the solid aggregate. \cite{katz2006dynamics} and \cite{rudge2015melt} showed that if this viscosity is (effectively) non-Newtonian with a power-law exponent of $\sim$6, the angle is reduced to match the observations. However, this exponent was measured by \cite{king2010stress} to be $\sim$$1.5\pm 0.3$ at 95\% confidence---almost Newtonian. Furthermore, these isotropic theories cannot explain the radial melt segregation in torsion experiments.  In contrast, a theory of anisotropic Coble creep with Newtonian viscosity can explain the radial segregation \citep{qi2015experimental}. This theory was derived by \cite{takei2009viscous} from grain-scale considerations of anisotropic solid contiguity under deviatoric stress. It predicts that viscous resistance to deformation is reduced in the direction of minimum contiguity. It also predicts the emergence of porosity bands and, if the contiguity tensor aligns with the principal stress directions, that the bands grow fastest at low angles, consistent with experiments \citep{takei2013consequences}.  However, in laboratory experiments that produce bands, the grain-scale contiguity is misaligned by about 15$^\circ$ \citep{qi2015experimental, qi2018influence}, which corresponds to a theoretical prediction of high-angle porosity bands (this discrepancy is resolved by better measurement of contiguity, according to \cite{seltzer2023melt}). Nonetheless, viscous-anisotropy theory gives rise to an effective dilatancy that we discuss in \S\ref{sec:anisotropic-viscosity}, below.

Another challenge is to explain the characteristic wavelength of the high-porosity bands observed in experiments.  All of the theories noted above lack mode selection; instead, they predict the rate of band growth to plateau at decreasing wavelength. Several studies have invoked processes driven by interfacial energy to regularise the growth-rate spectrum. \cite{bercovici2016mechanism} incorporated capillary effects in a diffuse-interface approximation of a sharp porosity interface \citep{sun2004diffuse}---however, sharp interfaces emerge in experiments only long after the onset of instability.  \cite{takei2009generalized} and \cite{king2011experimental} hypothesised that variation of surface tension drives dissolution/precipitation reactions. When coupled with chemical diffusion in the melt phase, these reactions damp instability growth at small wavelengths. 

The theory of dense granular suspensions, as reviewed by \cite{guazzelli2018rheology}, holds promise in providing a simple and unified explanation for all of these observed patterns. A central feature is the anisotropic compressive stress between solid particles caused by shearing flow \citep{bagnold1954experiments}. The coupling between shear and compression is the consequence of microphysical interaction of suspended particles \citep{brady1997microstructure}. This behaviour is demonstrated empirically in various studies, but \cite{deboeuf2009particle} provide a particularly fascinating example and discussion. If the suspended solid phase is not rigidly confined, it can undergo a net dilation due to shear \citep{reynolds1885lvii, boyer2011unifying}. 

In a suspension contained within a constant volume, net dilatancy is prohibited but the solid fraction can vary internally. \cite{besseling2010shear} shows that shearing, dense suspensions are susceptible to a banding instability; this instability is modelled in terms of a suspension viscosity that increases with solid fraction. Their results run parallel to the theory for band emergence in partially molten rock \citep{stevenson1989spontaneous} except that in a suspension, the growth rate of bands also depends on the dilatancy. Moreover, \cite{morris1999curvilinear} shows that for suspension flows in cylindrical geometry (i.e., pipe flow, parallel-plate or cone-and-plate torsion), radial segregation of liquid and solid phases is predicted, consistent with experiments. Again, there is a parallel with results for partially molten rock in torsion \citep{qi2015experimental, qi2018influence} and pipe-flow \citep{quintanilla2019radial} configurations. In all of these flowing suspensions, the dilatancy stress plays a central role.

Another aspect of granular physics that may be relevant here is non-local fluidity (inverse viscosity). This concept was developed in the context of emulsions \citep[e.g.,][]{goyon2008spatial, bocquet2009kinetic} and adapted to  granular suspensions \citep{kamrin2012nonlocal}. The theory states that the flow-response to stress at a point in the granular medium is sensitive to the fluidity in a neighbourhood around that point. This neighbourhood has a typical size, $\xi$, of order $10\times$ the grain size and decreasing with the square root of shear stress. \cite{henann2013predictive} demonstrate that simulated shear zones, forced by a spatial discontinuity in boundary velocity, are regularised by non-local fluidity to a width that is consistent with experiments.  Hence non-local fluidity appears promising in regularising the growth spectrum of shear bands in experiments on partially molten rock.

The fundamentally granular nature of partially molten rock and the relevant predictions from theories of dense granular suspensions motivate the present work. Our aims are to develop a theory for partially molten rock that incorporates granular dilatancy and non-local fluidity, and to compare predictions of that theory to the results of laboratory experiments. We note that in doing so, we are applying granular physics at solid fractions above what is typically considered the jamming fraction, at which the solid phase becomes immobile. However, in crystalline materials at high homologous temperatures, grain boundaries behave as a viscous fluid that allows grains to slide past each other \citep{ashby1972boundary}. In this context, the solid phase can still be mobilized if sufficient shear stress is applied \citep{heussinger2009jamming}.  Moreover, \textit{in-situ} observations of polycrystalline aggregates deforming at high solid fraction show a clear link between grain-boundary sliding and dilatancy \citep{walte2005deformation, kareh2017dilatancy}.  Hence, we assert that although the grains of partially molten rocks are not rigid, they nonetheless undergo grain-boundary sliding that is associated with a compressive intergranular stress and may lead to dilatancy.

Mechanical decreases in solid fraction, including by dilatancy, are here referred to as decompaction. The poro-viscous theory of partially molten rock relates the decompaction rate to the pressure difference between the liquid and solid phases in a viscous constitutive law \citep{mckenzie1984generation}.  This approach differs from suspension theory, in which the solid phase exerts zero resistance to changes in solid fraction \citep{guazzelli2011physical}. It also differs from theories for dry granular media, where the solid fraction is a decreasing function of the shear-strain rate \citep{forterre2008flows}, and from soil mechanics, where the solid fraction is predicted to evolve toward a critical state as a function of the total strain \citep{oda2020mechanics}. However, it seems that these isotropic dynamics may be incompletely understood. For example, \cite{kabla2009dilatancy} found empirical evidence that dilatancy and shear-independent compaction compete in the evolution of solid fraction. By combining poro-viscous decompaction with dilatancy stress, our theory may provide new insight in this regard.

Previous authors have incorporated granular dilatancy into discussions and models of geological materials, going back at least to \cite{mead1925geologic}. It has been invoked in crystal-rich deforming magma \citep[e.g.,][]{smith1997shear, petford2020igneous}, in lower-crustal shear zones \citep{menegon2015creep}, and in gouge-filled fault zones \citep[e.g.,][]{marone1990frictional, segall2010dilatant}. Dilation has been considered in competition with compaction \citep{paterson2001granular, niemeijer2007microphysical} and as a microphysical mechanism responsible for rate-and-state friction \citep{chen2016rate}. It may play a role in regulating glacial sliding \citep{warburton2023shear} and in a range of geomorphological processes \citep{jerolmack2019viewing}. Dilation is associated with Riedel shear zones \citep[e.g.,][]{dresen1991stress, bedford2021role}, which appear at the same angle as bands in partially molten rock. It might be expected that partially molten rock shares certain behaviour with other granular, geological materials. In the present work, we find that incorporation of granular dilatancy and non-local fluidity brings predictions of a poro-viscous compaction theory into quantitative agreement with experimental results. 

The paper is organised as follows.  In \S\ref{sec:experiments} we review torsion experiments on partially molten rocks and highlight their key results. We present our rheological model in \S\ref{sec:rheology}.  Then, in \S\ref{sec:solutions}, we provide the governing equations and analyse them in terms of radial segregation and band formation. This analysis is followed by quantitative comparison with experiments in \S\ref{sec:comparison} and a discussion in \S\ref{sec:discussion}.

\section{Laboratory experiments and key observations}\label{sec:experiments}

Previously described laboratory experiments provide a motivation and context for testing the theory developed here. We focus on experiments conducted on partially molten rock, typically synthesized from mixtures of $\sim$95\% olivine grains and $\sim$5\% mid-ocean ridge basalt, sometimes with a small percentage of chromite \citep[e.g.,][]{holtzman2003stress, king2010stress, qi2015experimental}. The olivine grains are polydisperse, typically with a mean diameter of $\sim$10~$\mu$m.  Samples are hydrostatically hot-pressed to remove gas-filled bubbles prior to deformation. After hot-pressing, they have a nominally uniform melt fraction, $\phi_0$.

\begin{figure}
    \centering
    \includegraphics[width=\textwidth]{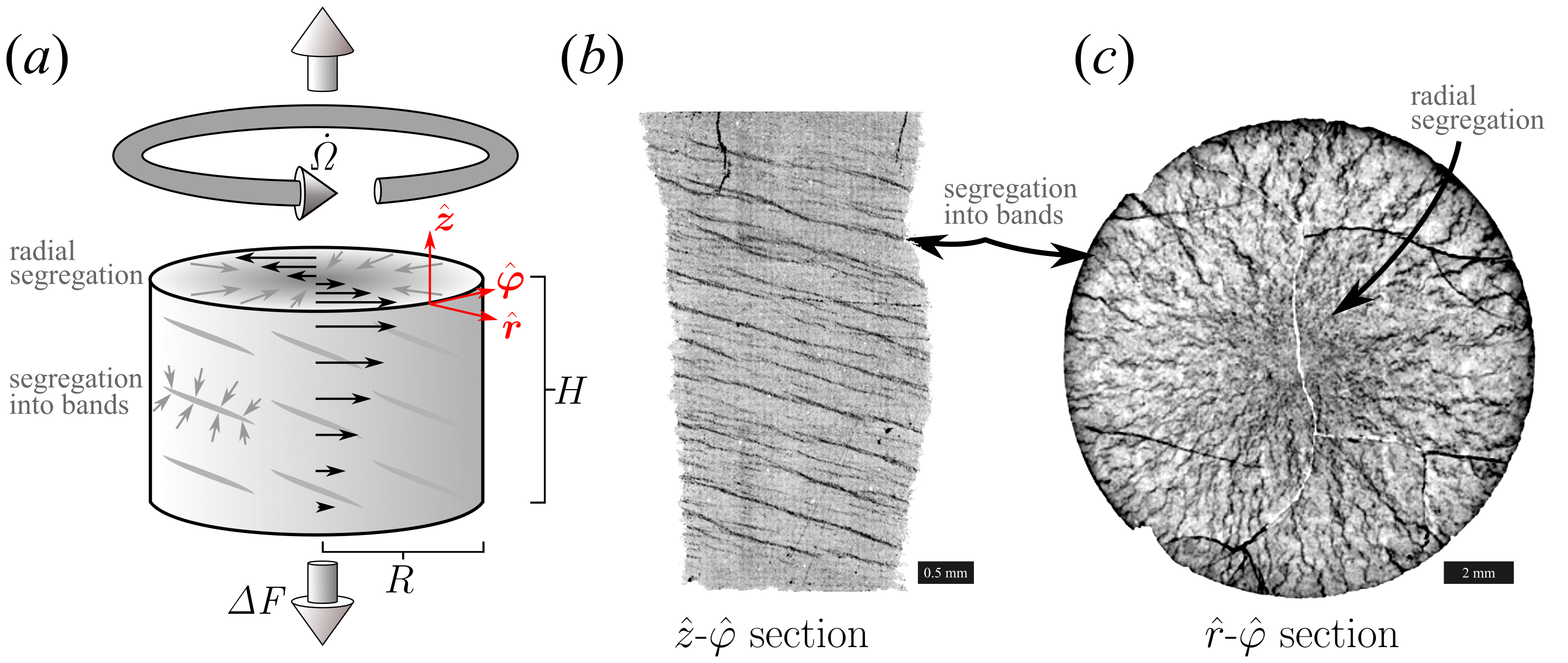}
    \caption{Experimental configuration and representative results. \textit{(a)} Schematic diagram of a deforming experimental sample and the emergent patterns of melt segregation. Experiments are conducted at high confining pressure and high temperature. After achieving a specified twist, the sample is quenched, sectioned, and polished to reveal the distribution of melt (solidified to glass) and crystalline, granular solid. \textit{(b)} A tangential section showing high-porosity bands (black) at low angle to the shear plane \citep[$\phi_0=0.04,\;\gamma=1.5$;][]{king2010stress}. \textit{(c)} A transverse section showing radially inward migration of melt \citep[$\phi_0=0.10,\;\gamma=5.0$;][]{qi2015experimental}. Cracks visible in panels (b) and (c) are a consequence of the rapid quench and decompression after deformation.}
    \label{fig:schematic}
\end{figure}

The experiments are conducted in a gas-medium, triaxial-deformation apparatus \citep{paterson1990rock} with a confining pressure of 300~MPa and temperatures of $\sim$1225\textdegree C. The samples are jacketed to separate them from the confining gas. Under these conditions, the basalt is molten and the olivine (and chromite) grains are solid. Torsional deformation is imposed on the sample, although some experiments have also been conducted in direct shear \citep{holtzman2003stress, holtzman2007stress}. The distribution of porosity within the sample is not measured \textit{in situ}. Rather, the experiment is quenched, sectioned, and imaged at high resolution to calculate the porosity field.

The essential characteristics of these torsional experiments are outlined in figure~\ref{fig:schematic}a. Cylindrical samples with height $H$ and outer radius $R$ are deformed by a circular platen that turns with angular velocity $\dot{\Omega}\zhat$ about the axis of the cylinder. At low strains, when the sample remains nominally uniform, the imposed twist induces an azimuthal velocity field $U(r,z)\phihat$ that is axisymmetric. The velocity component $U$ increases linearly in the $\zhat$ and $\rhat$ directions. On cylindrical surfaces, the deformation is approximately that of simple shear; the magnitude of the shear strain (and its rate) increase from zero at the twist axis to a maximum value at the outer boundary of the sample.

The outer boundary of the sample is sealed in an impermeable, nickel jacket. The radial normal stress at the jacket is maintained constant by the confining gas pressure. At the temperatures of the experiments, the viscosity of nickel is greater than that of the basaltic liquid and less than the granular olivine aggregate. This viscosity contrast enables the jacket to shear with negligible resistance, but discourages its intrusion into the pore space of the sample. Hence its effect on the sample falls somewhere between two limiting cases. In one limit, the jacket inhibits all radial flow at the boundary by isolating the volume of the sample. In the other limit, it transmits the full confining pressure into the pore space between olivine grains. There is empirical evidence that the reality is closer to the first of these limits, but the details have not been measured or quantified.

Two critical observations have arisen from torsion experiments on partially molten rocks. The first is the emergence of high-porosity sheets separated by compacted, low-porosity lenses after a shear strain of $\gamma \sim 1$ \citep{holtzman2003stress}. These sheets are usually measured in cross-section, as in figure~\ref{fig:schematic}b, and hence referred to as bands. They have a characteristic spacing and form at 15--20\textdegree{} to the shear plane \citep{holtzman2007stress}. This angle is similar to that of Riedel shear zones \citep{dresen1991stress}, but significantly lower than what would be expected if the bands were normal to the direction of maximum tension (45\textdegree). Furthermore, individual bands are embedded in a nominally simple-shear flow and hence with time, they are rotated to higher angles. However, despite this necessary rotation, the band-angle distribution remains roughly unchanged with increasing strain \citep{king2010stress}. 

The second critical observation is that with progressive twist, liquid melt segregates from the solid grains and migrates toward the center of the cylinder \citep{qi2018influence}.  This migration leads to azimuthally averaged porosity, measured over the transverse section shown in figure~\ref{fig:schematic}c, that decreases with radius. Experiments to increasing values of total twist (reported as shear strain $\gamma$ at the outer radius) exhibit greater segregation and a steeper radial porosity gradient \citep{qi2015experimental}.

The present study aims to explain these observations in terms of the physics of dense granular suspensions.
 
\section{Rheological model}\label{sec:rheology}

Our rheological model of a two-phase aggregate, comprising a contiguous matrix of solid grains and its melt-saturated, permeable pore-space, is based on the poro-viscous theory derived by \cite{mckenzie1984generation} and reviewed by \cite{katz2022dynamics}.  The melt is present with volume fraction $\phi$ (the porosity) that varies in space and time.  This variation is accommodated by (de)compaction of the solid matrix, but both phases are incompressible. To incorporate dilatancy effects, we take inspiration from theories of suspensions \citep{brady1997microstructure, fang2002flow, guazzelli2018rheology} and append a term that hypothetically quantifies the normal stresses generated by grain--grain interactions during shearing flow. The constitutive law for the effective stress is then 
\begin{equation}
    \label{eq:rheological-model}
    \boldsymbol{\sigma}^\text{eff} = \cmpvisc\cmp\identity 
    + 2\shearvisc\strrten - \dilvisc\boldsymbol{\Lambda}\strr,
\end{equation}
where $\cmpvisc$, $\shearvisc$ and $\dilvisc$ are dynamic viscosities for isotropic, deviatoric and dilational deformation, respectively, $\identity$ is the identity tensor, and where
\begin{equation}
    \label{eq:define-symbols-1}
    \cmp \equiv \Div\vel^s,\qquad
    \strrten \equiv \tfrac{1}{2}\left[\Grad\vel^s + \left(\Grad\vel^s\right)^T - \tfrac{2}{3}\cmp\identity\right],\qquad
    \boldsymbol{\Lambda}\equiv 
    \left(\begin{array}{ccc}
        1 & 0  & 0 \\
        0 & \Lambda_\perp & 0 \\
        0 & 0 & \Lambda_\times
    \end{array}\right),
\end{equation}
are the decompaction rate, deviatoric strain-rate tensor, and particle-stress anisotropy tensor, respectively. We have introduced $\vel^s$, the solid velocity field, and $\strr \equiv \sqrt{\strrten:\strrten/2}$ is the second invariant of the deviatoric strain-rate tensor. \cite{deboeuf2009particle} provides theoretical context, insightful commentary, and empirical justification for the dilatancy term in \eqref{eq:rheological-model}.

The particle-stress anisotropy tensor $\boldsymbol{\Lambda}$ is used to model the normal stresses generated by a particle-laden flow that is locally approximated as simple shear \citep{guazzelli2011physical, guazzelli2018rheology}.  It is written with reference to a coordinate system aligned with the simple shear.  The $\Lambda_{11}$ direction is taken to be the direction of flow (indicated by $\parallel$); the $\Lambda_{22}$ direction is normal to the shear plane (hence we denote it $\Lambda_\perp$); the $\Lambda_{33}$ direction is the direction of the vorticity vector (and hence denoted $\Lambda_\times$).   The $\Lambda_{11}$ entry is factored out and lumped with $\dilvisc$. Therefore, $\Lambda_\perp$ and $\Lambda_\times$ are dimensionless particle-normal-stress ratios. Previous work has shown that the values of these parameters may be constrained by comparison of model predictions with carefully designed experiments \citep{morris1999curvilinear, fang2002flow, guazzelli2018rheology}.  In the flow geometries considered below (Cartesian or cylindrical), the particle-stress anisotropy tensor can be straightforwardly aligned with the experimental deformation geometry; in general, it must be aligned with respect to the principal axes of the flow \citep{miller2009suspension}. 

The isotropic part of the effective stress is where dilatancy modifies the physics.  We see this by taking $-1/3$ the trace of the effective stress tensor in equation \eqref{eq:rheological-model},
\begin{equation}
    \label{eq:trace-rheological-model}
    \dilvisc\strr\, \trace{\boldsymbol{\Lambda}}/3 = \omp\left(P^s-P^\ell\right) + \cmpvisc\cmp,
\end{equation}
where $P^j = -\trace{\boldsymbol{\sigma}^j}/3$ is the pressure of phase $j$. This equation states that the shear-strain rate has two possible consequences for isotropic deformation. If $\cmpvisc=0$, there is no viscous resistance to compaction and shear generates a positive effective pressure. This is equivalent to suspension theory \citep[e.g.,][]{deboeuf2009particle}.  If, in contrast, there is zero effective pressure ($\Delta P=0$), then shear causes dilation. This has a parallel in soil mechanics, where the dilatancy angle $\psi$ gives the kinematic relationship between shear strain and dilation  \citep[e.g.,][]{oda2020mechanics}.  In the present context, we can compute the dilatancy angle as $\tan\psi \equiv \dilvisc \trace{\boldsymbol{\Lambda}}/3\cmpvisc$. The more general case, of interest here, is where both the effective pressure and the compaction viscosity are nonzero.

To complete the rheological model, we require expressions for the dependency of the three viscosities on melt fraction $\phi$. Empirical constraints and theoretical models of the shear viscosity $\shearvisc$ and compaction viscosity $\cmpvisc$ are summarised by \cite{katz2022physics}.  Shear viscosity has been measured over a range of melt fractions; \cite{kelemen1997review} showed that it is well-described by an exponential decrease with liquid fraction $\phi$.  Theory for Coble creep, where compaction is accommodated by diffusion of grain mass along grain boundaries and through the melt-filled pores, indicates that the compaction viscosity is a multiple of $\sim$5/3 larger than the shear viscosity \citep{takei2009viscous, rudge2018viscosities}.  There are no empirical measurements of the dilitation viscosity $\dilvisc$ of partially molten rock, nor are there are microstructural models.  Experiments on particle suspensions by \cite{deboeuf2009particle} show an exponential weakening of particle normal stress with liquid fraction.  On this basis, and for simplicity in the absence of further information, we take $\dilvisc$ to be an unknown multiple of $\shearvisc$.  Hence the viscosities are given by 
\begin{equation}
    \label{eq:viscosities-coble-creep}
    \shearvisc = \eta_0\text{e}^{-\lambda(\phi-\phi_0)},\qquad  \cmpvisc = 5\shearvisc/3,\qquad \dilvisc=D_0\shearvisc,
\end{equation}
where $\eta_0$ is a reference value of shear viscosity at reference melt fraction $\phi_0$, $\lambda \approx 27$ is the porosity-weakening factor, and $D_0$ is an unknown, dimensionless constant.

We obtain a constraint on $D_0$ by requiring positive entropy production under any combination of shear and isotropic deformation. The dissipation-rate density arising from \eqref{eq:rheological-model} is  
\begin{equation}
    \label{eq:diss-rate}
    \Psi = \cmpvisc\cmp^2 + 4\shearvisc\strr^2 - \dilvisc\strr\left(\dot{\varepsilon}_\parallel + \Lambda_\perp\dot{\varepsilon}_\perp + \Lambda_\times\dot{\varepsilon}_\times\right),
\end{equation}
where $\dot{\varepsilon}_\parallel,\, \dot{\varepsilon}_\perp,\, \dot{\varepsilon}_\times$ are the normal components of the strain-rate tensor in a coordinate system aligned with simple shear. Assuming isotropic dilatancy $\boldsymbol{\Lambda} = \identity$, equation~\eqref{eq:diss-rate} becomes $\Psi = \cmpvisc\cmp^2 + 4\shearvisc\strr^2 - \dilvisc\strr\cmp$, and into this we substitute the viscosities of equation~\eqref{eq:viscosities-coble-creep}.  We find that $\Psi$ is positive definite if $0 \le D_0 < 4\sqrt{5/3} \approx 5$ and therefore limit consideration to values of $D_0$ within this range. 

Finally, in combining our rheological model with conservation equations governing the flow, we consider the granular physics discussed by \cite{kamrin2012nonlocal}, which adopts a model for emulsions by \cite{goyon2008spatial}.  They show that macroscopic, irreversible shear is accommodated by grain-rearrangement events at the microscopic scale. In partially molten rock, geometric compatibility of the grain packing dictates that grains cannot rotate freely; their rotations must be compatible with those of neighbouring grains \citep{rudge2021micropolar}.  Hence deformation is necessarily dispersed by grain--grain interaction during rearrangement events.   This non-local interaction means that the viscosity at a point in the medium is influenced by the viscosity at points within a distance $\xi$, known as the cooperativity length. \cite{kamrin2012nonlocal} express this interaction in terms of a non-local fluidity---the inverse of the non-local shear viscosity $\shearviscnl$. We rewrite their fluidity equation in terms of a non-local viscosity,
\begin{equation}
    \label{eq:non-local-viscosity}
    \shearviscnl^{-1} = \shearvisc^{-1} + \xi^2\delsq \left(\shearviscnl^{-1}\right).
\end{equation}
Evidently, if $\xi=0$, then the non-local viscosity reduces to $\shearvisc$.  For $\xi>0$, this equation imposes a minimum scale of viscosity variation.  \cite{goyon2008spatial} measure $\xi$ as a function of $1-\phi$ and find that it increases to about 5$\times$ the grain diameter at a solid fraction of 85\%.  We shall see below in \S\ref{sec:simpleshear} that $\xi>0$ serves to regularise the spectrum of instability growth. 

\section{Analysis} \label{sec:solutions}

To explore the consequences of the hypothesised rheological model, we adopt the formulation of mass and momentum conservation for a partially molten rock deforming at zero Reynolds number \citep{mckenzie1984generation}. We make a Boussinesq approximation, taking density as constant for both phases, assume zero mass transfer between phases, and neglect gravitational body forces on the basis that they are much weaker than the shear tractions imposed in experiments. With these assumptions, the phase densities vanish from the equations.  The coupled system of conservation equations becomes
\begin{subequations}
    \label{eq:conservation-equations}
    \begin{align}
        \label{eq:compaction}
        \cmp &= \Div \mobility \Grad P^\ell,\\
        \label{eq:stokes}
        \Grad P^\ell  &= \Div2\shearviscnl\strrten + \Grad\cmpviscnl\cmp - \Div\dilviscnl\boldsymbol{\Lambda}\strr,\\
       \label{eq:solid-mass}
       \lagd{s}{\phi}{t} &= \omp\cmp.
    \end{align}
\end{subequations}
The first, known as the compaction equation, is obtained from Darcy's law by eliminating the liquid velocity using the two-phase continuity equation. It includes the fluid mobility $\mobility = M_0 (\phi/\phi_0)^n$, which represents the ratio of the porosity-dependent permeability and the constant liquid viscosity.  The second equation is a statement of force balance in the two-phase aggregate.  The third equation is mass conservation for the solid phase (porosity is transported by the solid velocity). These equations are standard \citep{katz2022dynamics}, except for two modifications. The first modification is use of the non-local viscosity $\shearviscnl$ in equation~\eqref{eq:stokes}, which couples it to equation~\eqref{eq:non-local-viscosity} governing the non-local viscosity. For consistency, we use the non-local viscosity in \eqref{eq:viscosities-coble-creep} to compute non-local compaction $\cmpviscnl$ and dilitation $\dilviscnl$ viscosities.  The second modification is the last term on the right-hand side of equation~\eqref{eq:stokes}, which captures the hypothesised dilatancy effects. The classical model is recovered for $\xi,D_0\to0$.

In the subsections below, we investigate the consequences of these two modifications.  We do so in the context of torsional deformation and boundary conditions that mimic the laboratory experiments described in section~\ref{sec:experiments}.  

\subsection{Radial segregation in parallel-plate torsion}\label{sec:torsion}

Torsional flow embeds simple shear into a cylindrical geometry with the potential for hoop stress.  The experiments described in section~\ref{sec:experiments} demonstrate that parallel-plate torsional flow drives solid radially outward and liquid radially inward.  This phenomenon is consistent with the behaviour of dense suspensions undergoing parallel-plate torsional flow \citep{merhi2005particle} but in contrast to the Poiseuille flow of appendix~\ref{sec:poiseuille}. We consider cone-and-plate torsional flow (where the plates are \textit{not} parallel) in appendix~\ref{sec:cone-plate-torsion}.

To understand the radially outward transport of solid grains in terms of dilatancy and particle-stress anisotropy, we work in a cylindrical geometry with coordinates ($r,\varphi,z$), as shown in figure~\ref{fig:schematic}a.  We consider a cylinder of partially molten rock with outer radius $R$, azimuthal symmetry in $\varphi$ and, instantaneously at $t=0$, with uniform porosity $\phi_0$.  At this instant, the solid flow is assumed to have zero $\zhat$ component, a fixed azimuthal component, and an unknown radial component. This flow is described by
\begin{equation}
    \label{eq:torsion-velocity-ansatz}
    \vel^s = V\rhat + \frac{\dot{\Omega}\,r z}{H}\phihat, \qquad\cmp=\frac{1}{r}\pd{}{r}rV,\qquad\strrten = \left(\begin{array}{ccc}
        \pd{V}{r} - \frac{\cmp}{3} & 0 & 0 \\
         0 & \frac{V}{r} - \frac{\cmp}{3} & \tfrac{\dot{\Omega r}}{2H} \\
         0 & \tfrac{\dot{\Omega r}}{2H} & -\frac{\cmp}{3}
    \end{array}\right),
\end{equation}
where $V(r)$ is the unknown radial component of the solid velocity field, $\dot{\Omega}$ is the constant twist rate and $H$ is the uniform gap between the parallel plates. We linearise the strain-rate intensity under the assumption that dilatancy is driven by the forced shear such that
\begin{equation}
    \label{eq:approx-invariant-pptorsion}
    \strr\sim\dot{\Omega}r/2H.
\end{equation}
This choice eliminates a feedback whereby the anisotropic part of the dilatancy drives additional dilatancy. While it may be physically reasonable, it will reduce the predicted dilatancy at a given value of $D_0$ relative to the case where the feedback is included.

We use $\strr$ in the radial component of force-balance equation \eqref{eq:stokes} to write
\begin{equation}
    \pd{P^\ell}{r} = 3\eta_0\pd{}{r}\frac{1}{r}\pd{}{r}rV - \frac{D_0\dot{\Omega}R^2}{6H}(2\Lambda_\times-1). 
\end{equation}
We then combine this with the compaction equation~\eqref{eq:compaction} to eliminate the liquid pressure and integrate once. Rescaling $r$ with the outer radius $R$ and $V$ with the characteristic scale
\begin{equation}
    \label{eq:torsion-vel-scaling}
    [V] = \frac{D_0\dot{\Omega}R^2}{6H},
\end{equation}
we obtain the dimensionless equation
\begin{equation}
    \label{eq:cylindrical-radial-ode-torsion}
    \pd{}{r}\frac{1}{r}\pd{}{r}rV - \frac{V}{\ndcl^2} = 2\Lambda_\times-1.
\end{equation}
Here we have introduced 
\begin{equation}
    \ndcl \equiv \frac{\sqrt{3\eta_0M_0}}{R},
\end{equation}
the ratio of the compaction length to the outer radius. The compaction length is an emergent length scale over which perturbations to the solid--liquid pressure difference are relaxed by decompaction \citep{mckenzie1984generation, spiegelman1993physics, katz2022dynamics}. 

The normal-stress difference on the right-hand side of equation \eqref{eq:cylindrical-radial-ode-torsion} arises from the particle-stress anisotropy tensor $\boldsymbol{\Lambda}$, with the coordinates aligned such that the flow direction is $\phihat$ and the vorticity direction is $\rhat$.  The boundary condition at the centre of the cylinder is $V(0)=0$.  With this constraint, equation~\eqref{eq:cylindrical-radial-ode-torsion} admits the solution
\begin{equation}
    \label{eq:cylindrical-radial-ode-gensol}
    V(r) = \pi\ndcl^2\left(\tfrac{1}{2}-\Lambda_\times\right)\left[A I_1(r/\ndcl) - L_1(r/\ndcl)\right],
\end{equation}
where $I_n(z)$ is the modified Bessel function of the first kind, $L_n(z)$ is the modified Struve function, and
$A$ is a constant to be determined by matching the boundary condition at the dimensionless outer radius $r=1$. Two end-member cases can be considered for this outer boundary condition.  

\subsubsection{Outer boundary condition: no normal flow} 
\begin{figure}
    \centering
    \includegraphics[width=\textwidth]{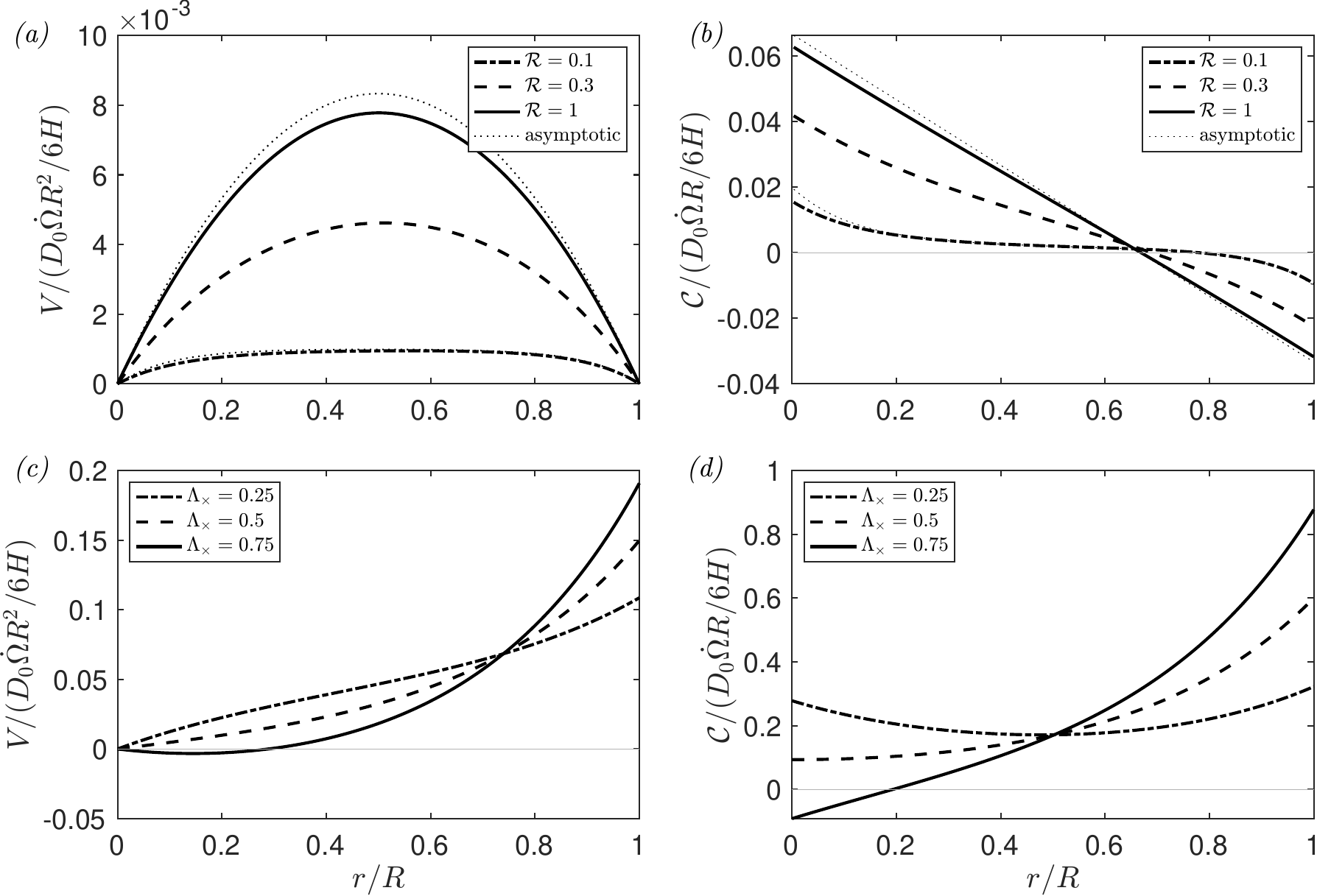}
    \caption{Parallel-plate torsion flow at $t=0$. Nondimensional solutions of eqn.~\eqref{eq:cylindrical-radial-ode-torsion} with uniform $\eta_\phi=\eta_0$. Top panels have outer boundary condition $V(R)=0$; bottom panels have zero-effective-stress outer boundary condition given in eqn.~\eqref{eq:confining-pressure-nd-bc}. \textit{(a)} Analytical solutions~\eqref{eq:torsion-zero-outer-analytical} with outer boundary condition $V(R)=0$ and with $\Lambda_\times=0.45$. In the limit of $\ndcl \gg 1$, the dimensionless solution is asymptotic to $V(r)\sim (2\Lambda_\times - 1) (r^2-r)/3$. In the other limit, $\ndcl\ll1$, the matched asymptotic solution is $V(r)\sim \ndcl^2(2\Lambda_\times - 1)\left[-1+\exp(-r/\ndcl) + \exp(-(1-r)/\ndcl))\right]$. \textit{(b)} Decompaction rate with $\Lambda_\times=0.45$. \textit{(c)} Analytical solution \eqref{eq:torsion-zeroeffstress-analytical} with outer boundary condition~\eqref{eq:confining-pressure-nd-bc} and with $\ndcl=0.3$. \textit{(d)} Decompaction rate with $\ndcl=0.3$.}
    \label{fig:parallel-torsion-radial-soln}
\end{figure}

In this case, a rigid outer cylinder requires that at $r=R$, the radial component of velocity is $V(R)=0$.  Then the analytical solution to dimensionless equation~\eqref{eq:cylindrical-radial-ode-torsion} is 
\begin{equation}
    \label{eq:torsion-zero-outer-analytical}
    V(r) = \pi\ndcl^2\left(\tfrac{1}{2} -\Lambda_\times\right)\left[ \frac{L_1(1/\ndcl)}{I_1(1/\ndcl)}I_1(r/\ndcl)-L_1(r/\ndcl)\right].
\end{equation}
This result demonstrates that the sign of $V(r)$ is determined by the size of $\Lambda_\times$.  In figure~\ref{fig:parallel-torsion-radial-soln}(a), we have chosen $\Lambda_\times = 0.45$ such that $V(r)>0$. This choice is qualitatively consistent with experimental results (\S\ref{sec:experiments}) where the solid is observed to move outward with progressive twist. Evidently, for the outer boundary condition $V(1)=0$, any choice of $\Lambda_\times$ satisfying
\begin{equation}
    \label{eq:parallel-torsion-constraint}
    \Lambda_\times < 1/2
\end{equation}
is also qualitatively consistent. For this range of $\Lambda_\times$, the hoop stress generated by dilatancy in the flow direction is stronger than the dilatant normal stress in the radial (vorticity) direction. As noted by \citet{takei2013consequences}, a compressive hoop stress drives solid radially outward.

\subsubsection{Outer boundary condition: no normal effective stress} Alternatively, we can consider the case where the partially molten cylinder is surrounded by an inviscid fluid, held at a dimensional confining pressure $P_c$.  This pressure must be balanced by the phase-averaged traction at the boundary and therefore,  $\rhat\cdot\overline{\boldsymbol{\sigma}}\cdot\rhat = -P_c$. Assuming that the liquid pressure is continuous at $r=R$, we obtain the boundary condition
\begin{equation}
    \rhat\cdot \boldsymbol{\sigma}^\text{eff} \cdot\rhat = 0\quad\text{at}\quad r=R.
\end{equation}
Expanding this condition using \eqref{eq:rheological-model} and the approximate second invariant \eqref{eq:approx-invariant-pptorsion}, then non-dimensionalising $r$ with $R$ and $V$ with $[V]$, we obtain the dimensionless boundary condition
\begin{equation}
    \label{eq:confining-pressure-nd-bc}
    \frac{V}{3} + \pd{V}{r} = \Lambda_\times \quad\text{at}\quad r=1.
\end{equation}
This condition yields a different value of $A$ and the general solution \eqref{eq:cylindrical-radial-ode-gensol} becomes 
\begin{equation}
    \label{eq:torsion-zeroeffstress-analytical}
    V(r) = \pi\ndcl^2\left(\tfrac{1}{2}-\Lambda_\times\right)
    \left[\frac{3L_0\left(\tfrac{1}{\ndcl}\right) - 2\ndcl L_1\left(\tfrac{1}{\ndcl}\right)+\tfrac{3}{\pi\ndcl}\frac{\Lambda_\times}{1/2-\Lambda_\times}}{3I_0\left(\tfrac{1}{\ndcl}\right) - 2\ndcl I_1\left(\tfrac{1}{\ndcl}\right)}I_1\left(\tfrac{r}{\ndcl}\right)-L_1\left(\tfrac{r}{\ndcl}\right)\right].
\end{equation}
This function is plotted in figure~\ref{fig:parallel-torsion-radial-soln}(c)--(d); the curves are computed with $\ndcl=0.3$ and values of $\Lambda_\times$ that span $1/2$. The radial component of solid velocity $V$ is generally positive, indicating outward solid flow and decompaction across all radii. This outward flow is again driven by the compressive hoop stress. Distinct from the rigid outer boundary condition, however, condition~\eqref{eq:confining-pressure-nd-bc} allows the solid to move outward at the outer boundary.  Hence in this case, torsion with any $\Lambda_\times$ causes the solid cylinder to expand radially, imbibing liquid across the outer boundary.

The pattern of flow in figure~\ref{fig:parallel-torsion-radial-soln}(c) is slightly different for $\Lambda_\times=0.75$, where there is a region of $V<0$ at inner radii.  The inner part of this region is associated with compaction $\cmp<0$, as shown by the solid curve in panel~(d).  The driving force is again dilatancy, but in this case with $\Lambda_\times>1/2$, the radial dilatant normal stress plays a significant role.  As is the case for Poiseuille flow in appendix~\ref{sec:poiseuille}, faster shear at larger radii drives solid inward.  But with zero radial effective stress at the outer boundary, dilatancy also drives outward solid flow, radial expansion of the cylinder, and radial imbibition of liquid.

\subsubsection{Outward force on pistons due to dilatancy}

The results depicted in figure~\ref{fig:parallel-torsion-radial-soln} for both outer boundary conditions are valid instantaneously at $t=0$, when all properties are uniform with radius.  At this initial instant, we compute an axial force outward on the plates, parallel to $\zhat$, that arises from dilatancy. This calculation provides a prediction to be compared with laboratory measurements. Details of the calculation are in appendix~\ref{sec:force-calculation}.  The main result is that the axial force is dominated by the direct effect of dilatancy in the flow-perpendicular direction, and hence scales as 
\begin{equation}
    \label{eq:excess-force-scaling}
    \Delta F \approx \torque D_0 \Lambda_\perp / 3R,
\end{equation}
where $\torque$ is the torque that causes a twist-rate of $\dot\Omega$ at $t=0$.  $\Delta F$ is the outward force in excess of that due to the confining pressure surrounding the sample. Using typical laboratory values for $\torque$ and $R$ in \eqref{eq:excess-force-scaling} (appendix~\ref{sec:force-calculation}), we find that the excess force is on the order 10\% of the force due to the typical confining pressure. 

In detail, the excess force $\Delta F$ can deviate from the simple prediction of \eqref{eq:excess-force-scaling}. Figure~\ref{fig:force} in appendix~\ref{sec:force-calculation} plots this deviation for both outer-boundary-condition cases over a range of $\ndcl$ and $\Lambda_\times$.  When the outer boundary is closed to solid flow ($V(R)=0$), the excess force is close to the simple scaling above; when the outer boundary has zero effective stress (eqn.~\eqref{eq:confining-pressure-nd-bc}), dilatancy in the radial direction leads to a net decompaction of the sample and a reduction in the excess axial force by approximately one half.

\subsubsection{Finite time and steady state}
The analysis of parallel-plate torsion to this point has considered the instantaneous problem at $t=0$, when the domain is uniform in porosity.  The instantaneous flow requires that this uniform state is subsequently lost by radial segregation of solid and liquid (and, as shown empirically and below in \S\ref{sec:simpleshear}, by a banding instability).  Appendix~\ref{sec:torsion-finite-time} derives the system of equations, simplified from \eqref{eq:conservation-equations}, that governs the finite-time evolution of the radial distribution of porosity.  The non-uniform porosity $\phi(r)$ leads to a radial dependence of mobility $\mobility$ and aggregate viscosity $\shearviscnl$.

\begin{figure}
    \centering
    \includegraphics[width=\textwidth]{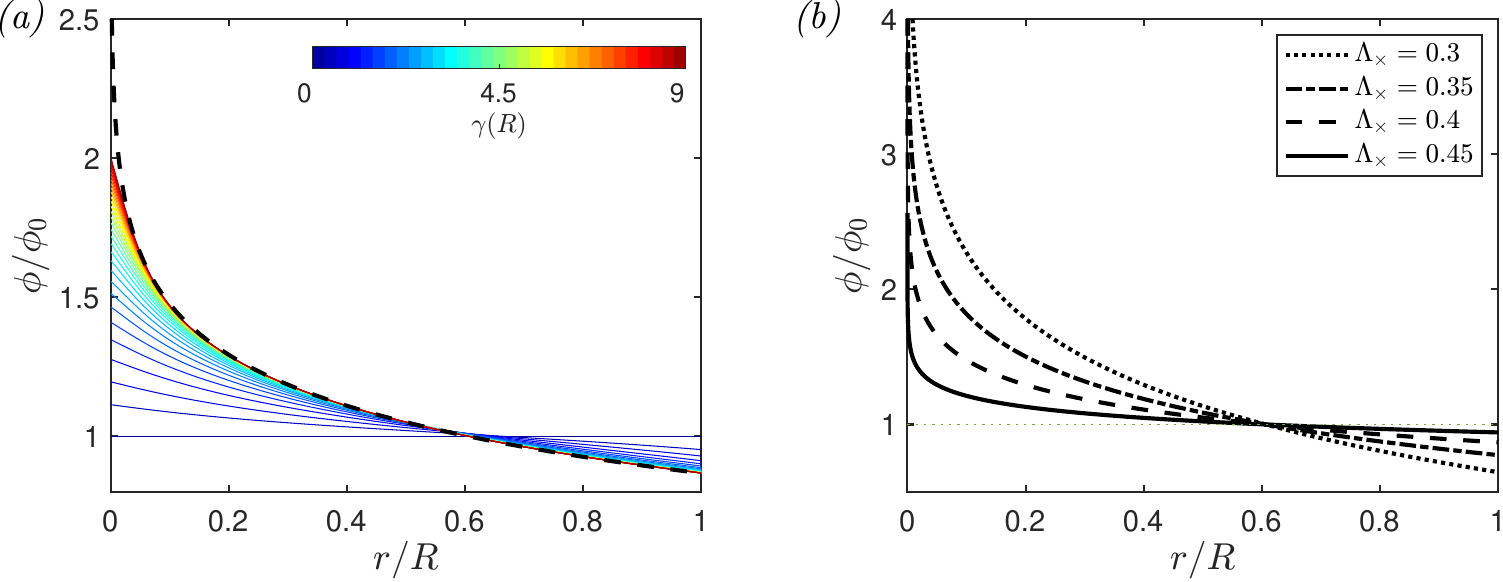}
    \caption{Parallel-plate torsion at $t\ge0$ and $t\to\infty$. Coloured curves show the time-dependent, numerical solution to the system~\eqref{eq:radtor-system} for porosity $\phi(r,t)/\phi_0$. Black curves show the analytical, steady-state solution~\eqref{eq:torsion-ss-analytical} for $\xi=0$. Both panels use empirically motivated values $\lambda=27$ and $\phi_0=0.07$. \textit{(a)} Solutions with $\Lambda_\times=0.4$, $\ndcl=0.3$ and $\xi/R=0.03$ at various outer-radius strains $\gamma(R)$. \textit{(b)} Steady solutions for four values of $\Lambda_\times$.}
    \label{fig:torsion-finitestrain-ss}
\end{figure}

A series of time-dependent numerical solutions are plotted in figure~\ref{fig:torsion-finitestrain-ss}a, coloured according to the shear strain $\gamma$ at the outer radius of the domain $R$. Details of the numerical method are in appendix~\ref{sec:torsion-finite-time}; code is available in an online repository \citep{katz2023software}. This calculation uses $\ndcl=0.3$ and $\Lambda_\times=0.4$ with boundary condition $V(R)=0$. At the smallest finite strains, the porosity distribution has the shape of the $t=0$ solution for $\cmp(r)$, shown in fig.~\ref{fig:parallel-torsion-radial-soln}b. With increasing strain (time), the porosity contrast between the centre and outer radius increases.  However, the evolution slows and ceases as the porosity distribution approaches a steady state.

In the steady state and with $\xi=0$, the radial component of the solid velocity is zero and radial force-balance equation~\eqref{eq:stokes} reduces to $\rhat\cdot\left(\Div \dilvisc\boldsymbol{\Lambda}\strr\right) = 0$ or, after expanding and rearranging,
\begin{equation}
    \label{eq:torsion-steady-balance}
    \dilvisc = \Lambda_\times\left(2\dilvisc + r\dilvisc'\pd{\phi}{r}\right),
\end{equation}
where $\dilvisc'$ is the derivative of the dilation viscosity with respect to its argument, $\phi$.  In solutions to this equation, a steady state is reached when the force of the hoop stress (left-hand side) balances the force of the radial normal stresses (right-hand side). The compressive hoop force, associated with a coefficient of unity in $\boldsymbol{\Lambda}$, is due to azimuthal dilatancy that pushes solid radially outward. The radial normal force, associated with the coefficient $\Lambda_\times$ in $\boldsymbol{\Lambda}$, has two causes: first, the gradient in radial dilatancy due to the torsional shear ($\strr\propto r$), and second, the gradient in radial dilatancy due to the steady-state radial gradient in porosity.

Laboratory experiments that impose torsional deformation, discussed in \S\ref{sec:experiments}, have an azimuthally averaged porosity that decreases with radius. On the basis of the predicted compaction rate at $t=0$, shown in figure~\ref{fig:cone-plate-torsion-radial-soln}, we can infer that this porosity structure is consistent with the no-normal-flow boundary condition and $\Lambda_\times < 1/2$. It is unclear whether these experiments approach a steady-state radial porosity profile, or would do so at larger strains. If a steady state can be achieved, then equation~\eqref{eq:torsion-steady-balance} requires that $\dilvisc'<0$ independent of the specific form of $\dilvisc$; in other words, it requires that the dilatancy stress at fixed strain rate decreases as porosity increases. 

In equation~\eqref{eq:viscosities-coble-creep} we specified that $\dilvisc$ has the form $\dilvisc=D_0\shearvisc$. Given the exponential dependence of $\shearvisc$ on $\phi$, it follows that $\dilvisc' = -\lambda\dilvisc$. With this we can solve \eqref{eq:torsion-steady-balance} to give
\begin{equation}
    \label{eq:torsion-ss-analytical}
    \phi(r,t\to\infty) = \phi_0 - \frac{1/2-\Lambda_\times}{\Lambda_\times\lambda/2}\left(\ln r + \frac{1}{2}\right),
\end{equation}
where $r$ has been non-dimensionalised with the outer radius $R$ and we have used global conservation of liquid mass to determine the constant of integration (see appendix~\ref{sec:torsion-finite-time} for details). This function is plotted as black curves in figure~\ref{fig:torsion-finitestrain-ss} for various values of $\Lambda_\times$; in  \S\ref{sec:comparison} we compare it with measurements from laboratory experiments. The logarithmic singularity in \eqref{eq:torsion-ss-analytical} for $r \to 0$ is removed by non-local viscosity when cooperativity length $\xi>0$.  This is evident in figure~\ref{fig:torsion-finitestrain-ss}a by comparison between the numerical solution at late time (red curve; $\xi=0.1\delta$) and the steady solution (black curve; $\xi=0$). 

\subsection{Simple shear between parallel plates} \label{sec:simpleshear}

\begin{figure}
    \centering
    \includegraphics[width=\textwidth]{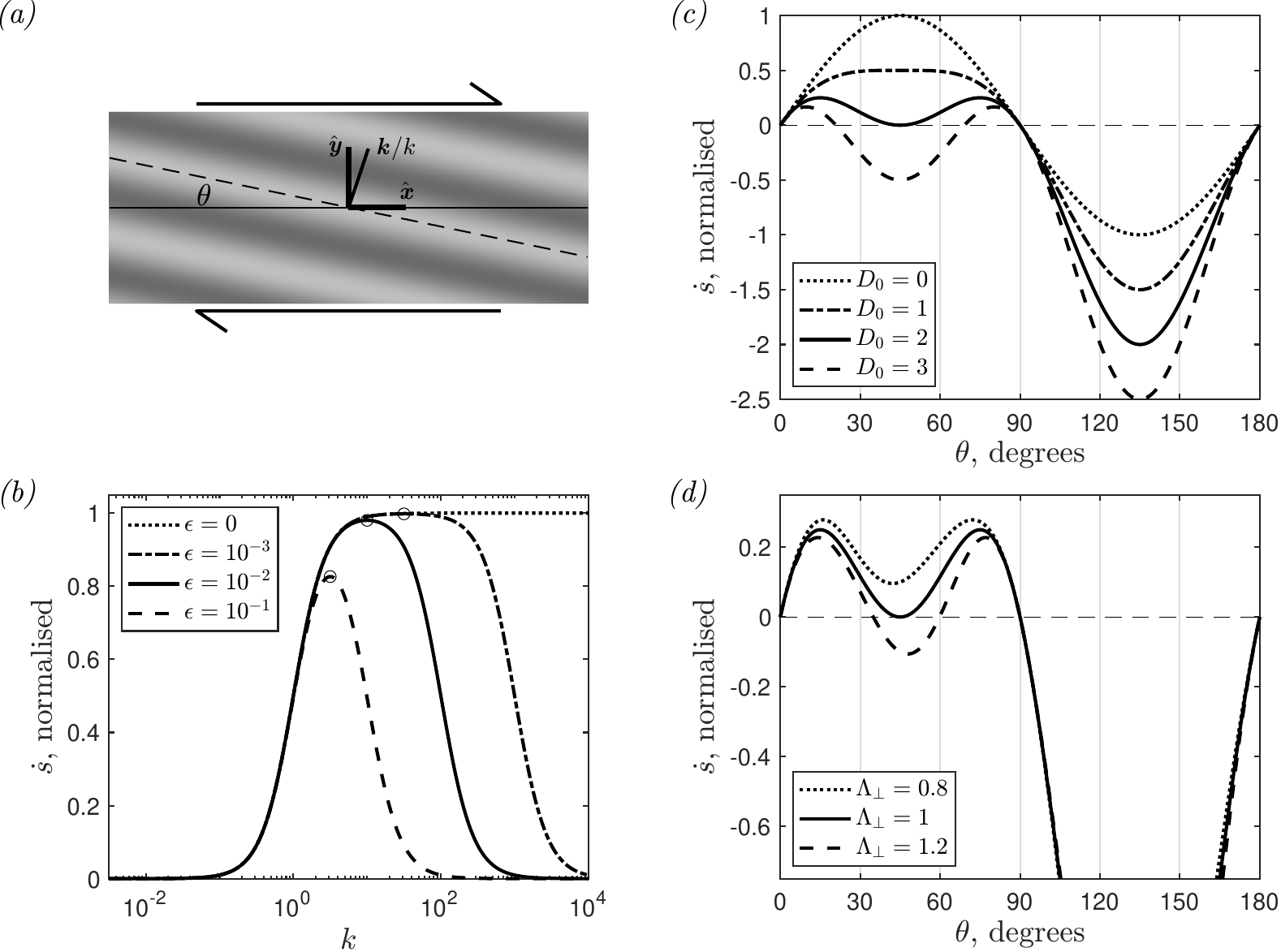}
    \caption{Growth rate of sinusoidal perturbations under a simple-shear flow from eqn.~\eqref{eq:simpleshear-growthrate}. \textit{(a)} Schematic diagram showing a finite region of the infinite domain. Grayscale shows the perturbed porosity field. \textit{(b)} Growth rate as a function of wavenumber $k$ for $D_0=0$ at $\theta=45^\circ$. Circles represent the growth rate computed at $k^*=\epsilon^{-1/2}$. \textit{(c)} Growth rate angular factor as a function of $\theta$ with $\Lambda_\perp = 1$, and values of $D_0$ given in the legend. \textit{(d)} Growth rate angular factor as a function of $\theta$ with $D_0=2$, and values of $\Lambda_\perp$ given in legend.}
    \label{fig:simpleshear}
\end{figure}

Here, following the analysis of \cite{spiegelman2003linear}, we investigate the stability of a two-dimensional simple-shear flow with initial melt fraction $\phi_0 + \phi_1(\boldsymbol{x},t)$, where $\vert\phi_1\vert\ll\phi_0$ is a perturbation. A schematic diagram is shown in figure~\ref{fig:simpleshear}(a).  The coordinate system is oriented such that $\xhat$ is in the flow direction and $\yhat$ is in the direction perpendicular to the shear plane.  We assume invariance in the $\zhat$ (vorticity) direction and take the $x$--$y$ plane to be infinite; hence there is no need to impose boundary conditions.  The procedure is a standard linearised stability analysis, detailed in \citet[][Chap.~7]{katz2022dynamics} and sketched in the next paragraph. \cite{alisic2016torsion} provides a three-dimensional analysis for torsion in cylindrical coordinates, but this adds mathematical complexity without additional physical insight.

We use equation~\eqref{eq:stokes}  to eliminate the pressure gradient from \eqref{eq:compaction} and obtain an equation governing the irrotational part of the velocity field. Then we take the curl of equation~\eqref{eq:stokes} to obtain an equation governing the solenoidal part. These are coupled to equations \eqref{eq:non-local-viscosity} and \eqref{eq:solid-mass} for viscosity and solid mass.  We expand variables into a steady, background state and a time-dependent perturbation that is arbitrarily small at $t=0$.  The perturbations are assumed to be proportional to $\exp[i\boldsymbol{k}(t)\cdot\boldsymbol{x} + {s}(t)]$, where $\boldsymbol{k}(t)$ is a time-dependent wave vector that changes direction and magnitude with the background flow. After linearising the governing equations, we solve the leading-order balance for the base state.  This is a simple-shear flow with velocity gradient $\dot\gamma$ and zero compaction rate.  Using the base-state solution, we solve the perturbation equations to obtain the dimensionless growth rate $\dot{s}$ as a function of dimensionless wave-vector magnitude $k$ and wavefront angle to the shear plane $\theta \equiv \tan^{-1}k_x/k_y$.  In the present case, we obtain
\begin{equation}
    \label{eq:simpleshear-growthrate}
    \dot{s} = (1-\phi_0)\frac{\lambda}{3}\frac{k^2}{(1+k^2)(1+\epsilon^2 k^2)}\left[\sin 2\theta - \frac{D_0}{2}\frac{\left(\sin^2\theta + \Lambda_\perp\cos^2\theta\right)\sin^2 2\theta}{1-\tfrac{D_0}{4}(1-\Lambda_\perp)\sin 4\theta}\right].
\end{equation}
In this equation, $\dot{s}$ has been made dimensionless by scaling with the background rate of shear $\dot{\gamma}$. Wavenumber $k$ has been made dimensionless by scaling with the compaction length $\delta \equiv \sqrt{3 M_0 \eta_0}$. We have introduced the ratio $\epsilon \equiv \xi/\delta$ representing the dimensionless cooperativity scale. 
Localisation phenomena in partially molten rock can emerge at scales smaller than the compaction length. In this case, localisation occurs because positive perturbations have lower viscosity, and hence decompact under resolved tension \citep{stevenson1989spontaneous,spiegelman2003linear}.  We refer to these perturbations as `bands,' making explicit reference to the high-porosity bands seen in experimental cross-sections. Below we discuss the dependence of band growth rate on wavenumber, angle and physical parameters.

Figure~\ref{fig:simpleshear}(b) shows the wavenumber dependence of the growth rate for several values of $\epsilon$ (assuming optimal orientation, $\theta = \theta^*$).  The curve for $\epsilon=0$ is the case with zero cooperativity of the viscosity field.  It shows the classical result that all wavelengths smaller than the compaction length ($k\gg1$) grow equally fast \citep{stevenson1989spontaneous}.  The use of the non-local viscosity with finite $\epsilon$ regularises the spectrum, imposing a short-wavelength cutoff at a dimensional wavelength $\sim \xi$, the cooperativity scale of the non-local viscosity.  Growth-rate curves in fig.~\ref{fig:simpleshear}(b) have a maximum at a dimensional wavenumber
\begin{equation}
    \label{eq:peak-growthrate-wavenumber}
    k^* = 1/\sqrt{\xi\delta}.
\end{equation}

Figure~\ref{fig:simpleshear}(c) shows the growth rate as a function of the angle $\theta$ between wavefronts and the shear plane.  Four curves show different values of the dilatancy viscosity prefactor $D_0 \ge 0$ (assuming $\boldsymbol{\Lambda} = \identity$, i.e., isotropic dilatancy).  The curve for $D_0=0$ corresponds to the case with no dilatancy, as studied by \cite{spiegelman2003linear}.  This case has positive growth rates between zero and 90$^\circ$, a range over which bands are subject to tension, and negative rates for angles greater than 90$^\circ$, which are subject to compression. The maximum growth rate occurs at $\theta^*=45^\circ$, where band wavefronts are perpendicular to the principal tension axis.  

For larger $D_0$ in fig.~\ref{fig:simpleshear}(c), dilatancy leads to peak growth rate at low and high angles.  In particular, the two maxima of growth rate occur at angles $\theta^*_{1,2}$ that vary with $D_0>1$, 
\begin{equation}
    {\theta}^*_1 = \arcsin(1/D_0)/2,\qquad {\theta}^*_2 = \pi/2 - \arcsin(1/D_0)/2. 
\end{equation}
A growth-rate peak at $\theta^*=15^\circ$, which is roughly that observed in experiments, corresponds to $D_0=2$.

Dilatancy has two competing effects that combine to produce this spectral shift with increasing $D_0$.  First, due to $\dilvisc'<0$, the background simple-shear flow causes a dilatancy perturbation that is exactly anti-phase with the porosity perturbation. This causes perturbations to decay at a rate that is independent of band angle $\theta$.  Second, the porosity perturbations create variations in shear viscosity, which in turn create perturbations in the rate of shear strain.  These drive variations in dilatancy that are exactly in-phase with variations in porosity; they hence contribute to perturbation growth. Critically, however, the growth rate associated with this second mechanism depends on band angle $\theta$.  Shear localises when bands are at low or high angle to the shear plane and therefore this effect is proportional to $\cos^22\theta$.  The combination of the two contributions of isotropic dilatancy is negative, overall, and proportional to $\sin^22\theta$.

Dilatancy in partially molten rock may be anisotropic, however. Experiments on dense granular suspensions, reviewed by \citet{guazzelli2018rheology}, have obtained inconclusive estimates of $\Lambda_\perp$, with some reporting $\Lambda_\perp$ increasing from unity with solid fraction, some reporting the opposite, and others reporting $\Lambda_\perp$ within error of unity for all solid fractions.  In figure~\ref{fig:simpleshear}(d) we compare growth-rate curves as a function of band angle for $\Lambda_\perp = 0.8,\,1,\,1.2$ for $D_0=2$.  The differences in the growth-rate peaks are subtle---likely indistinguishable on the basis of measured band-angle histograms from experiments.

\section{Comparison with laboratory data}\label{sec:comparison}

\begin{figure}
    \centering
    \includegraphics[width=\textwidth]{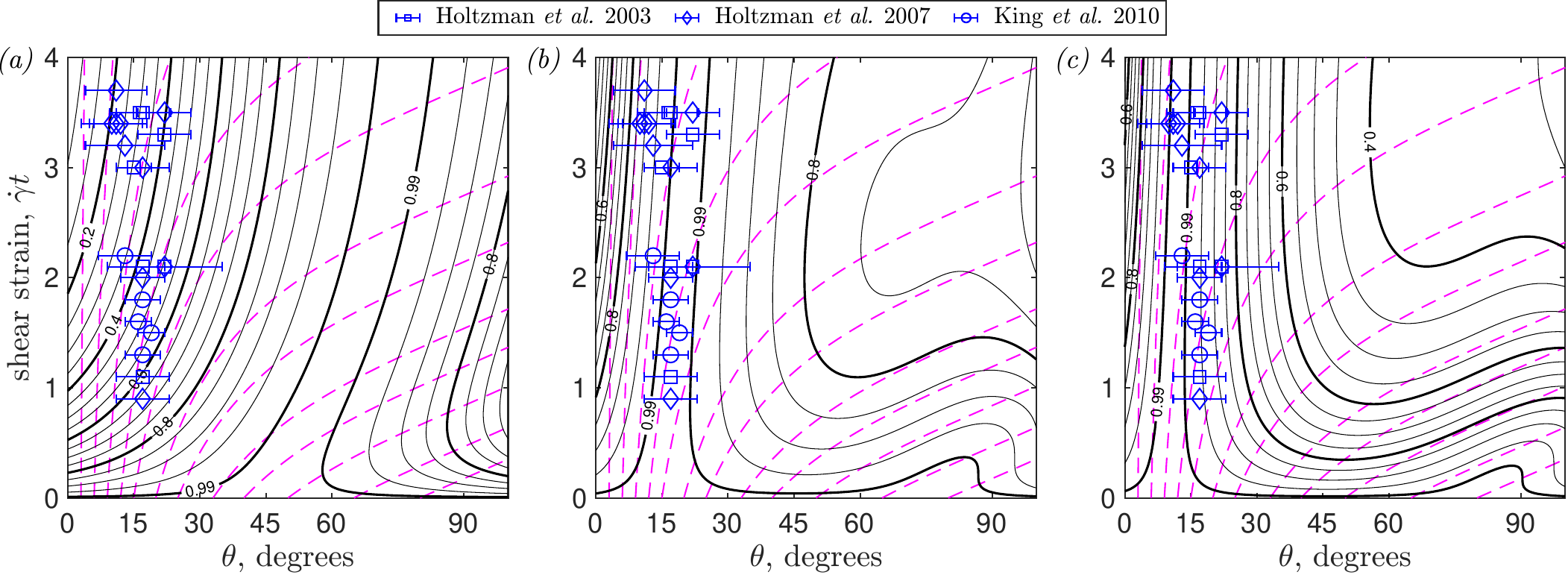}
    \caption{Angle spectra of porosity-band amplitude as a function of shear strain $\dot{\gamma}t$. The data points are the same in each panel. They record the mean angle from band-angle histograms of individual, published experiments (see legend); error bars are one standard deviation of the histogram.  Solid lines are contours of the band amplitude $\exp[s(t)]$, normalised over angles at each increment of strain. Dashed lines are passive advection trajectories (see text). Amplitude is computed by quadrature of the growth rate ${\dot{s}(t)}$ from equation~\eqref{eq:simpleshear-growthrate} with $\Lambda_\perp = 1$ and dimensionless $k(t=0) = k^* = \epsilon^{-1/2}$. Each panel has a different magnitude of dilatancy: \textit{(a)} $D_0=0$; \textit{(b)} $D_0=2$; \textit{(c)} $D_0=3$.}
    \label{fig:finitestrain}
\end{figure}

Published results from laboratory experiments (\S\ref{sec:experiments}) provide an opportunity to test our theory.  We begin by considering the best-established outcome of experiments, the localisation of melt into high-porosity bands. In particular, we first consider the angle that the bands make to the shear plane.  In figure~\ref{fig:finitestrain}, blue points with error bars indicate the mean and standard deviation of band angles from experiments quenched at shear strains between about 1 and 4. Each panel presents the same data set. The points form a coherent array with mean angles in the range 15--20$^\circ$, except at strains greater than about 3, at which the points spread out into a range of 10--25$^\circ$. The dashed lines, also identical in each panel, indicate trajectories that band angles would follow if rotated passively in the simple-shear flow \citep{katz2006dynamics}. 

Comparison of the array of blue points with the adjacent dashed lines clearly demonstrates that the evolution cannot be characterised as passive rotation of an initial set of bands. The maintenance of low angles over large strains has been attributed to successive generations of low-angle bands that draw melt from (and hence replace) previous generations as they undergo rotation to angles unfavourable for growth \citep{holtzman2005viscous,katz2006dynamics}. This process occurs at a finite perturbation amplitude and, strictly speaking, should not be described by solutions of linearised governing equations.  However, if the angular spectrum of growth rate (i.e., fig.~\ref{fig:simpleshear}c) remains approximately independent of strain, then a forward integration of $\dot{s}$ with respect to time (strain) along the trajectories of passive rotation might approximate the evolving angular spectrum of amplitude.  In this context, normalisation of the amplitude spectrum at each increment of strain might qualitatively represent melt redistribution. 

The contours of this finite-strain, normalised perturbation amplitude spectrum are plotted in figure~\ref{fig:finitestrain} for three different values of $D_0$ (each with $\boldsymbol{\Lambda} = \identity$). In panel~(a), for $D_0=0$, we see that without dilatancy, peak predicted amplitudes are far from observations.  In panel~(b), for $D_0=2$, and in panel~(c), for $D_0=3$, we see that peak predicted amplitudes occur at angles close to measured values.  The $D_0=2$ case is in better alignment at lower strains where nonlinear effects might be less important, and hence we take this value of the dilatancy pre-factor to be most appropriate in the context of the present model assumptions.

\begin{figure}
    \centering
    \includegraphics[width=0.5\textwidth]{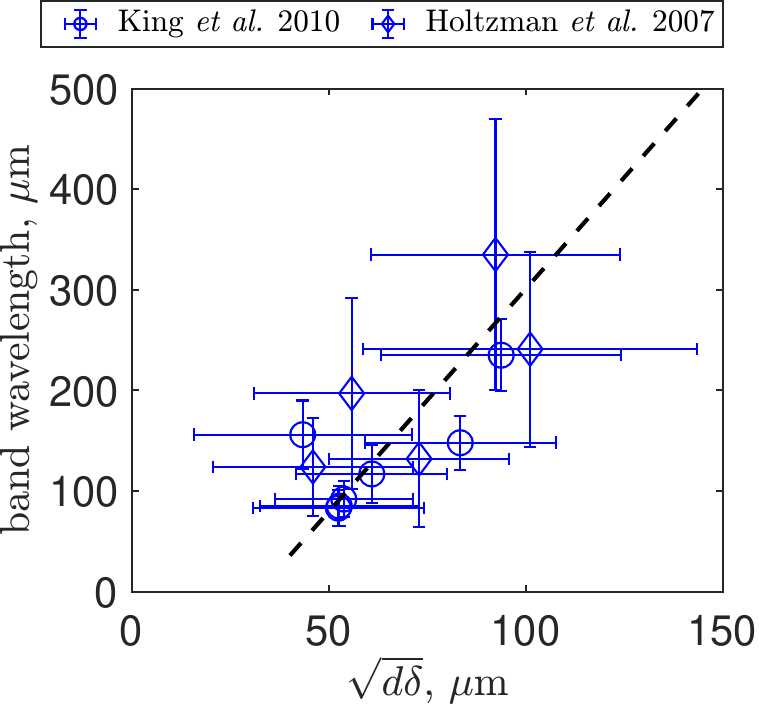}
    \caption{Wavelength of porosity bands in laboratory experiments (see legend) plotted against the geometric mean of the grain size $d$ and the compaction length. The data-source publications provide mean estimates of band spacing, band width, grain size, and compaction length. Band wavelength is calculated as the sum of mean band spacing and mean band width. Errors on measurements are propagated to give the error bars. The dashed line is a fit to the data respecting uncertainties on both axes \citep{york2004unified,weins23}.}
    \label{fig:wavelength_data}
\end{figure}

The linearised theory for porosity-band growth also provides a prediction of the wavelength with the largest growth rate.  Again, this prediction is strictly valid only near the onset of instability when the porosity contrast remains small.  Laboratory experiments that produce bands provide the opportunity to measure mean band width and spacing; summing these provides an estimate of the band wavelength.  However, as with band angles, these measurements are made in a nonlinear regime when the porosity contrast is large. So a comparison with theory is not strictly valid.  However, as with band angles, if the growth-rate spectrum remains roughly independent of strain, then a correspondence between theory and experiment might be expected even at larger strains. In that case, we would expect the observed wavelength to be proportional to $1/k^* = \sqrt{\xi\delta}$, where $\xi$ is the cooperativity length scale and $\delta$ is the compaction length. 

Figure~\ref{fig:wavelength_data} suggests that this correspondence may hold.  The blue symbols represent the wavelength of bands from experiments as a function of the geometric mean of the empirically known grain diameter $d$ and compaction length $\delta$. The cooperativity scale $\xi$ is understood to be proportional to grain diameter \citep{henann2013predictive} and hence $1/k^* \propto \sqrt{d\delta}$.  Although there is considerable uncertainty on the experimental estimates (particularly $\delta$), a linear trend is compatible with the data.

\begin{figure}
    \centering
    \includegraphics[width=0.6\textwidth]{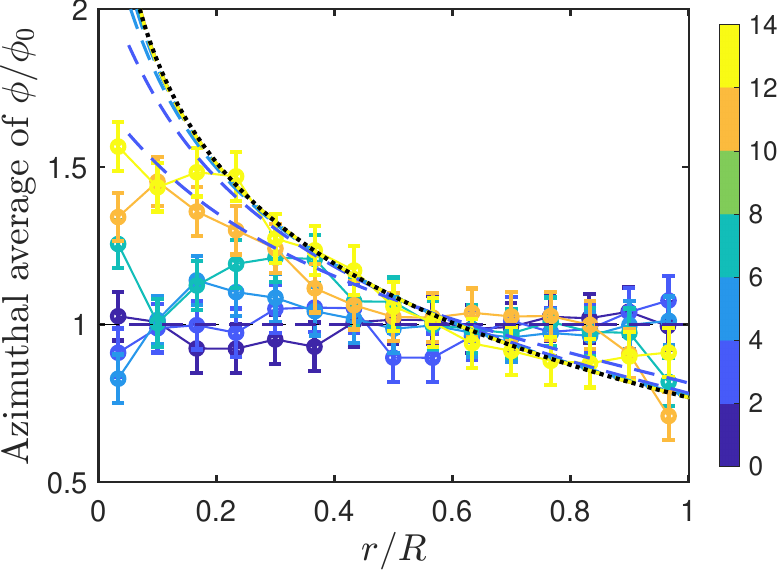}
    \caption{Radial distribution of porosity $\phi(r)$ normalised by the initial porosity $\phi_0$ in experiments and theory. Symbols represent porosity from laboratory experiments obtained by reprocessing high-resolution scans of transverse sections \citep{qi2015experimental,qi2018influence}; values are averages over rings of equal radial span. Error bars show the standard deviation of porosity in the undeformed ($\gamma=0$) experiment. Colours represent the shear strain at the outer radius $\gamma(R)$. The black, dotted line represents the steady-state solution \eqref{eq:torsion-steady-balance} with $\Lambda_\times=0.4$, $\phi_0 = 0.04$, $\lambda=27$. Dashed curves are numerical solutions to the system \eqref{eq:radtor-system} with the same parameters as for the steady curve and also $D_0=2$, $\ndcl=0.3$. The numerical solution is plotted at finite values of $\gamma(R)$, as given by their colour. Appendix~\ref{sec:torsion-finite-time} gives details of the numerical method.}
    \label{fig:torsion_chaodata}
\end{figure}

Finally, we evaluate model predictions of the radial distribution of porosity in torsion experiments quenched at different strains.  The experiments are a subset of those published by \cite{qi2015experimental} and \cite{qi2018influence}; we select only those in which the liquid phase is basaltic melt. We re-analyse high-resolution binary images of transverse sections, following the authors' published protocol, but averaging azimuthally over fewer, wider rings.  Recalculated porosities are presented as a function of normalized sample radius in figure~\ref{fig:torsion_chaodata}. The data are colored by the magnitude of shear strain at the outer radius, which ranges from zero to $\sim$14.  Three experiments with strains between 5 and 6 are averaged to produce one radial series.  Error bars represent the standard deviation of the averaged and normalised porosity at $t=0$, at which the porosity should be uniform. 

A qualitative conclusion can be immediately drawn by examining the data.  The boundary condition imposing zero effective stress at $r=R$ is not consistent with the experiments. This is because of the pattern of decompaction shown in figure~\ref{fig:parallel-torsion-radial-soln}d, which has the most rapid decompaction at the outer boundary. In contrast, the empirical data uniformly exhibit reduced porosity due to compaction there.  Hence we proceed using only the $V(R)=0$ condition of no radial flow at the outer boundary.

Model predictions are overlayed onto the laboratory data in figure~\ref{fig:torsion_chaodata}.  The black dotted line is the steady-state solution from equation~\eqref{eq:torsion-ss-analytical}.  This is computed with $\Lambda_\times = 0.4$, a value chosen to give an approximate match with the highest-strain experiment. (Note, however, that we have no evidence that the empirical porosity distribution at $\gamma\approx14$ is in steady state.) The dashed curves are numerical solutions of the time-dependent model (appendix~\ref{sec:torsion-finite-time}), plotted at values of outer-radius strain indicated by the colour of the curve. The shape of the model curves is in qualitative agreement with the data trends, within error. However, model porosity evolves more rapidly as a function of strain than the porosity in experiments.

There are two main difficulties in interpreting this mismatch in terms of parameter values or deficiencies of the theory.  The first is that we do not know whether the porosity distributions from experiments shown in fig.~\ref{fig:torsion_chaodata} are approaching a steady state, as predicted by the theory.  The uncertainties in the porosity and the coarse sampling in total strain make any inferences speculative.  Second, supposing that the experiments are evolving toward a steady state, we do not have a reliable estimate of the porosity distribution in that state. If we had constraints on that distribution (and knowledge of $\phi_0$ and $\lambda$), we could use equation~\eqref{eq:torsion-ss-analytical} to infer $\Lambda_\times$.

Advancing speculatively, we assume that the experiments do approach a steady state. We note that smaller $\Lambda_\times$ corresponds to larger $\infd\phi/\infd r$ at steady state (fig.~\ref{fig:torsion-finitestrain-ss}b). Assuming that our estimates for $\phi_0\approx 0.04$ and $\lambda\approx 27$ are sufficiently accurate, the data in figure~\ref{fig:torsion_chaodata} are indicative of $\Lambda_\times \lesssim 0.4$. We can estimate the timescale of porosity adjustment to steady state in the numerical solutions shown in figure~\ref{fig:torsion_chaodata}. Referring to the porosity evolution equation~\eqref{eq:solid-mass}, we approximate $\text{D}_s\phi/\text{D}t$ by $\Delta\phi/\tau$, where $\tau$ is the timescale over which porosity changes by $\Delta\phi$ from its initial value $\phi_0$ to its steady value at some given radius. According to \eqref{eq:solid-mass}, this change is driven by decompaction at a rate $\omp\cmp$; we approximate this rate as $\breve{\cmp}$, which we take to be the decompaction rate at $r=0,\,t=0$ (c.f.~fig.~\ref{fig:parallel-torsion-radial-soln}b). This rate is obtained by calculating $\cmp(r)$ at $t=0$ from the analytical solution~\eqref{eq:torsion-zero-outer-analytical} for radial velocity, re-dimensionalising, and evaluating at $r=0$ with $\ndcl \sim 1$. We then form the outer-radius strain at time $\tau$ as $\gamma(\tau) = \tau/\dot{\gamma} = \Delta\phi/\dot{\gamma}\breve{\cmp}$, where $\dot{\gamma} = \dot{\Omega}R/H$ is the outer-radius strain rate.  This obtains
\begin{equation}
    \label{eq:strain-scale-for-steady-state}
    \gamma(\tau) \sim \frac{18}{D_0\Lambda_\times\lambda},
\end{equation}
where we have used \eqref{eq:torsion-ss-analytical} to evaluate $\Delta\phi$ at $r = 1/\text{e}^2$.  For the parameters used in figure~\ref{fig:torsion_chaodata}, this gives an outer-radius strain of $\gamma\sim 1$ at which the simulated porosity has evolved to within a factor of $1/\text{e}$ of its steady value.  This estimate is comparable to the numerical solution of fig.~\ref{fig:torsion_chaodata}, but it is smaller than the empirical timescale by a factor of 10.  We cannot bring these timescales into agreement by changing $D_0$ because this is constrained by the angle of porosity bands (fig.~\ref{fig:finitestrain}), and we have already assumed that we know $\lambda$.  Reducing $\Lambda_\times$ thus appears to be an option.  According to figure~\ref{fig:torsion-finitestrain-ss}b, reducing $\Lambda_\times$ by a factor of 2 predicts a steady-state, radial porosity gradient much larger than observed at the largest empirically attained strain $\gamma$.  Experiments to larger $\gamma$ are needed to test this prediction. Alternatively, the discrepancy in timescales may be a consequence of nonlinear interaction of radial segregation with emergence of high-porosity layers, which is not captured in our models.

\section{Discussion}\label{sec:discussion}

We hypothesised that granular physics (i.e., dilatancy and non-local fluidity) shapes the patterns that emerge when partially molten rock is deformed in laboratory experiments. Our model predictions, based on a rheological formulation combining theories for poro-viscous compaction and dense granular suspensions, can be made quantitatively consistent with most aspects of the empirical data. This consistency arises through four key choices. First is the choice of a rigid boundary condition at the outer radius of the cylinder. Second is the choice of a reduced dilatancy in the vorticity direction of the particle-stress anisotropy tensor ($\Lambda_\times < 1/2$).  Together, these enable the prediction of radially inward melt segregation and compaction at outer radii, both of which are observed in experiments. The third choice is for $D_0\approx 2$, which predicts the emergence of porosity bands at 15--20\textdegree{} to the shear plane, as observed in experiments. The fourth choice is for a finite cooperativity length $\xi$, which regularises the growth-rate spectrum.

These choices are neither physically implausible nor empirically unreasonable. Therefore we assert that laboratory experiments provide support for our hypothesis, under the conditions (i.e., the strain rate) at which they are conducted. How (indeed, if) our theory extrapolates to the much slower strain rates under natural conditions depends on the physical processes that are occurring at the grain scale.  The grain-scale physics is discussed below, after we consider the relationship of our theory with that of anisotropic viscosity.

\subsection{Relationship to anisotropic viscosity}
\label{sec:anisotropic-viscosity}

A theory for anisotropic viscosity \citep{takei2009viscous} is also capable of explaining band angles and radial segregation \citep{takei2013consequences}. The basis for this theory is a model of Coble creep, where melt provides a fast pathway for circum-grain mass diffusion. In small-strain experiments \citep{takei2010stress}, deviatoric stress causes the melt to preferentially coat grain boundaries that have normal vectors in the direction of maximum tension. The theory predicts that this anisotropy in solid-phase contiguity causes an anisotropic creep response to deviatoric stress: the deviatoric-compression direction has higher viscosity than the deviatoric-tension direction. This anisotropy gives rise to an effective dilatancy that drives radial segregation and band angle \citep{qi2015experimental, takei2015consequences}, as also obtained here.

It may therefore make sense to think of the present, direct formulation of dilatancy as an effective description of underlying physics that is more fully described by anisotropic, Coble-creep viscosity.   However, there are reasons to doubt this view.  The first is that the bands that emerge in experiments on hot, partially molten rock are similar to Riedel shear zones in cold granular media \citep{schmocker2003granular, bedford2021role}, suggesting a common mechanism. Although both are cases of deforming granular media, Coble creep is thermally activated and does not contribute at low temperature.  A second reason is that the solid-phase contiguity tensor, measured in band-producing experiments, is not suitably oriented to predict low-angle bands in viscous anisotropy theory \citep{takei2013consequences, qi2015experimental} \citep[however, see][]{seltzer2023melt}. And a third reason is that viscous anisotropy theory has no inherent mechanism to regularise the band growth-rate spectrum. So it may be that the granular-medium hypothesis considered here represents a distinct physical mechanism, albeit with similar implications for observable features. 

Further work is required to develop experiments and analyses that can distinguish between these competing hypotheses.  For example, granular physics should be tested against the hysteresis measured in oscillating stress experiments \citep{takei2010stress}. To capture this behaviour may require extension of the present theory to include a fabric tensor \citep{mehrabadi1982statistical} and its evolution.  A more fundamental approach, however, is to develop grain-scale models that consistently integrate a set of plausible physical processes. This could clarify the conditions under which grain-boundary sliding is accommodated by dilatancy and/or Coble creep.

\subsection{Physics at the grain scale} \label{sec:grain-scale-physics-discussion}

The creeping deformation of polycrystalline aggregates is known to occur by several grain-scale mechanisms: deformation of grains by the motion of lattice dislocations, shape-change of grains by mass diffusion down gradients of chemical potential induced by deviatoric stress, and grain-boundary sliding whereby the centre of mass of adjacent grains moves relative to one-another.  This latter mechanism is typical of athermal granular flow, in which an aggregate of rigid grains is required to (locally) dilate to accommodate the geometric incompatibilities of relative motion. At higher effective stress, geometric incompatibility might instead be resolved by cataclasis or by shape-change of grains through diffusion or dislocations. A sub-set of these mechanisms may simultaneously contribute to macroscopic deformation of the aggregate.

There is currently no unified model to predict the relative contributions of different mechanisms across a broad range of conditions. However, three basic expectations are relevant. First, low effective stress relative to the driving shear stress will favour dilatancy over shape-change of grains. Second, higher resistance to grain--grain sliding along a grain boundary (whether viscous or associated with a frictional yield stress) will favour shape change over sliding. And third, higher homologous temperatures will favour thermally activated processes of mass diffusion and dislocation motion over cataclasis. 

Saturation of the dense granular medium with a mobile, incompressible liquid may also be an important factor. (For this discussion, we consider the presence of liquid as being independent of homologous temperature even though, in the case of partial melt, the two are linked.) A greater volume fraction of liquid means that less geometric incompatibility is incurred by relative motion of grains, and hence grain-boundary sliding is promoted. Furthermore, a greater liquid pressure relative to the mean compressive stress of the solid framework (i.e., a low effective stress) also promotes sliding. However, there are two other points to consider.  First, in a poro-viscous context, non-zero effective stress causes (de)compaction, even in the absence of shear. Second, liquid transport over a finite distance through a porous medium occurs on a timescale and with a resistance controlled by the ratio of liquid viscosity to permeability.  This transport is required to accommodate (de)compaction and is therefore a control on the evolution of the effective pressure.

These considerations become important in the context of dilatancy within a sealed domain of fixed volume. If the enclosed, saturated granular medium is undergoing shear, then throughout the volume there is a compressive solid stress arising from grain interactions.  Dilation can occur locally within the volume, but only if it is balanced by compaction elsewhere. There are two relevant cases. If the strain rate is nominally uniform within the domain, then dilating and compacting regions emerge by instability on length and time scales that are set internally, as in \S\ref{sec:simpleshear}. Alternatively, if the strain rate has an imposed gradient, then this will organise the spatial pattern of dilation and compaction, as in \S\ref{sec:torsion} and appendices \ref{sec:cone-plate-torsion} and \ref{sec:poiseuille}. The particle-stress anisotropy is of fundamental importance in this latter case.

\subsection{Implications for natural systems}

In the shallow mantle near mid-ocean ridges, huge volumes of partially molten rock undergo deformation. These regions bear some similarity to the laboratory experiments considered here, but the strain rates are orders of magnitude smaller. Because the melt-filled pore network is vast and isolated, the effective pressure $\Delta P$ of the solid phase may be low.  Does shear cause dilatancy in this natural system?  We consider this by rearranging equation~\eqref{eq:trace-rheological-model} with $\Delta P\approx 0$,
\begin{equation}
    \frac{\cmp}{\strr} \sim \frac{\dilvisc}{\cmpvisc},
\end{equation}
where in this simple relationship, $\cmp\ge 0$ is entirely due to dilatancy. On the right-hand side is a ratio of the dilation and compaction viscosities. Our results here suggest this ratio is ${O}(1)$ for experiments; it might be much smaller in the mantle.  Hence the question of dilatancy in a natural system may reduce to understanding how $\dilvisc$ and $\cmpvisc$ properties differ in the natural system from in the laboratory.  How they scale with temperature, grain size and porosity, and whether they have a (non-linear) dependence on strain rate are questions to be resolved by future laboratory experiments and grain-scale physical models.

At depths shallower than the mantle, crustal magmatic systems can have larger strain rates when crystal-laden melt is injected into dikes and sills \citep{rivalta2015review}. These flows have lower solid fractions, at or below jamming, and hence more closely resemble granular suspensions \citep{smith1997shear, petford2020igneous}. Dilatancy should therefore be expected, and may drive crystals away from the walls of magma-filled fractures.  This would reduce the effective viscosity of the magma and promote propagation.

At shallower depths and lower temperatures in the crust, seismogenic faults may experience the effects of dilatancy. The slip across faults is often accommodated by rupture of asperities or shear of a granular medium called gouge. In both cases, a compressive stress will arise \citep{chen2016rate}. If dilation of the pore-space is possible, either by expansion of the contained air or inflow of water, there may be a decrease in friction and an unstable acceleration of slip.  In contrast, if dilation is prohibited, the compressive stress may increase friction and promote stable sliding.  In either case, once sliding has ceased, viscous creep may lead to slow compaction of the gouge (or asperities) to increase contact area and harden the fault. In this case, the compaction viscosity would control the timescale for frictional state evolution \citep{chen2017microphysically, thom2023microphysical}. 

Indeed, the viscous constitutive law~\eqref{eq:trace-rheological-model} relating compaction, dilatancy and the interphase pressure difference may be the most significant novelty of the present paper. This formulation bridges soil mechanics, suspension theory and theories for granular media with deformable grains. It may be broadly relevant in systems where deformation can occur on long time scales by irreversible creep. Such systems may be more common than is widely appreciated because slow granular processes have, until recently, gone largely unnoticed \citep[e.g.,][]{deshpande2021perpetual, houssais2021athermal}. 

\appendix

\section{Force on the plates in parallel-plate torsion} \label{sec:force-calculation}

We consider a torsion cell with radius $R$ and height $H$, aligned with a cylindrical coordinate system ($r,\varphi,z$). The bottom plate is fixed at $z=0$ and the top plate has angular velocity $\dot{\Omega}\zhat$.  At the instant $t=0$, the porosity is assumed to be uniformly  $\phi_0$ and the shear viscosity is uniformly $\eta_0$.  The normal force on the top plate is given by 
\begin{equation}
    F = -\int_0^R\int_0^{2\pi}\zhat\cdot\overline{\boldsymbol{\sigma}}\cdot\zhat\,r \, \infd\varphi\,\infd r
    = 2\pi\int_0^R(P^\ell - \zhat\cdot\boldsymbol{\sigma}^\text{eff}\cdot\zhat)\,r\,\infd r,
\end{equation}
where the negative sign gives the compression force (with a tension-positive sign convention for stress) and we have used $\boldsymbol{\sigma}^\text{eff} = \overline{\boldsymbol{\sigma}} + P^\ell\identity$.  Using equations \eqref{eq:rheological-model}, \eqref{eq:define-symbols-1} and \eqref{eq:viscosities-coble-creep} this becomes
\begin{equation}
    \label{eq:force-integral}
    F = 2\pi\int_0^R\left[P^\ell + \eta_0\left(D_0\Lambda_\perp\strr - \cmp\right)\right]r\,\infd r.
\end{equation}
For parallel-plate torsion, we assumed 
\begin{equation}
    \vel^s = V(r)\rhat + \dot{\Omega}\frac{rz}{H}\phihat
\end{equation}
and approximated $\strr\sim\dot{\Omega}r/2H$.  These are valid for a cylinder of finite height if the radial shear stress on the plates is zero. Using equations~\eqref{eq:compaction} and \eqref{eq:define-symbols-1} we can write
\begin{align}
    \label{eq:pressure-integral}
    P^\ell(r) &= \frac{1}{M_0}\int_0^rV(r')\infd r' + P^\ell(0),\nonumber\\
    &= P^\ell(R) - \frac{1}{M_0}\int_r^R V(r')\infd r'.
\end{align}
We recall that $M_0$ is the ratio of permeability to melt viscosity at $\phi=\phi_0$ and we assume that $P^\ell(R) = P_c$, the confining pressure of the experiment. Using~\eqref{eq:pressure-integral} and \eqref{eq:define-symbols-1} in \eqref{eq:force-integral} we obtain
\begin{equation}
    \label{eq:force-dimensional-0}
    F - F_c = 2\pi\int_0^R\left[-\frac{r}{M_0}\int_r^R V(r')\infd r' + \eta_0\left(\frac{D_0\Lambda_\perp\dot{\Omega}r^2}{2H} - \pd{}{r}rV\right)\right]\infd r,
\end{equation}
where $F_c \equiv \pi R^2 P_c$ is the force due to the confining pressure around the cylinder.

Integrating the second and third terms in \eqref{eq:force-dimensional-0} and non-dimensionalising $r$ with $R$ and $V$ with $[V] = D_0\dot{\Omega}R^2(2\Lambda_\times-1)/6H$ we obtain
\begin{equation}
    \label{eq:dimensional-force-integral-general}
    \Delta F = \frac{\torque D_0 \Lambda_\perp}{3R} \left[ 1 + \frac{2\Lambda_\times - 1}{\Lambda_\perp}\left(\frac{3\pi}{2}\mathcal{I}-V(1)\right)\right],
\end{equation}
where $\torque \equiv \pi\eta_0 R^4 \dot{\Omega}/H$ is the torque exerted to twist the sample and 
\begin{equation}
    \label{eq:pressure-integral2}
    \mathcal{I} \equiv -\frac{2}{\pi\ndcl^2}\int_0^1\int_r^1 V(r')\,\infd r'\,r\,\infd r
\end{equation}
is a dimensionless integral of dimensionless quantities. We emphasise that all three terms in the square brackets of equation~\eqref{eq:dimensional-force-integral-general} are dimensionless, but the factor outside the brackets is dimensional with units of force.

The first term in \eqref{eq:dimensional-force-integral-general} represents the direct effect of dilatancy; the second term represents the liquid pressure acting on the plate; the third term is the indirect effect of dilatancy, which causes a net decompaction of the cylinder.

To evaluate the second and third terms in \eqref{eq:dimensional-force-integral-general}, it remains to specify $V(r)$.  We consider two cases with different boundary conditions at the outer edge of the cylinder.

\subsection{No radial flow at $r=R$}
\label{sec:force-case-A}

For the condition $\rhat\cdot \vel^s(R)=0$ and uniform porosity (e.g., at $t=0$), we obtained the dimensionless solution given in equation~\eqref{eq:torsion-zero-outer-analytical}. For this case we have $V(0)=V(1)=0$ and hence the third term of \eqref{eq:dimensional-force-integral-general} gives zero contribution.  We use the solution \eqref{eq:torsion-zero-outer-analytical} for dimensionless $V$ in \eqref{eq:pressure-integral} to obtain
\begin{equation}
    \mathcal{I} = \int_0^1\int_r^1 \left[ \frac{L_1(1/\ndcl)}{I_1(1/\ndcl)} I_1(r'/\ndcl) - L_1(r'/\ndcl)\right]\infd r'\,r\,\infd r.
\end{equation}
This integral is evaluated by quadrature.

\begin{figure}
    \centering
    \includegraphics[width=\textwidth]{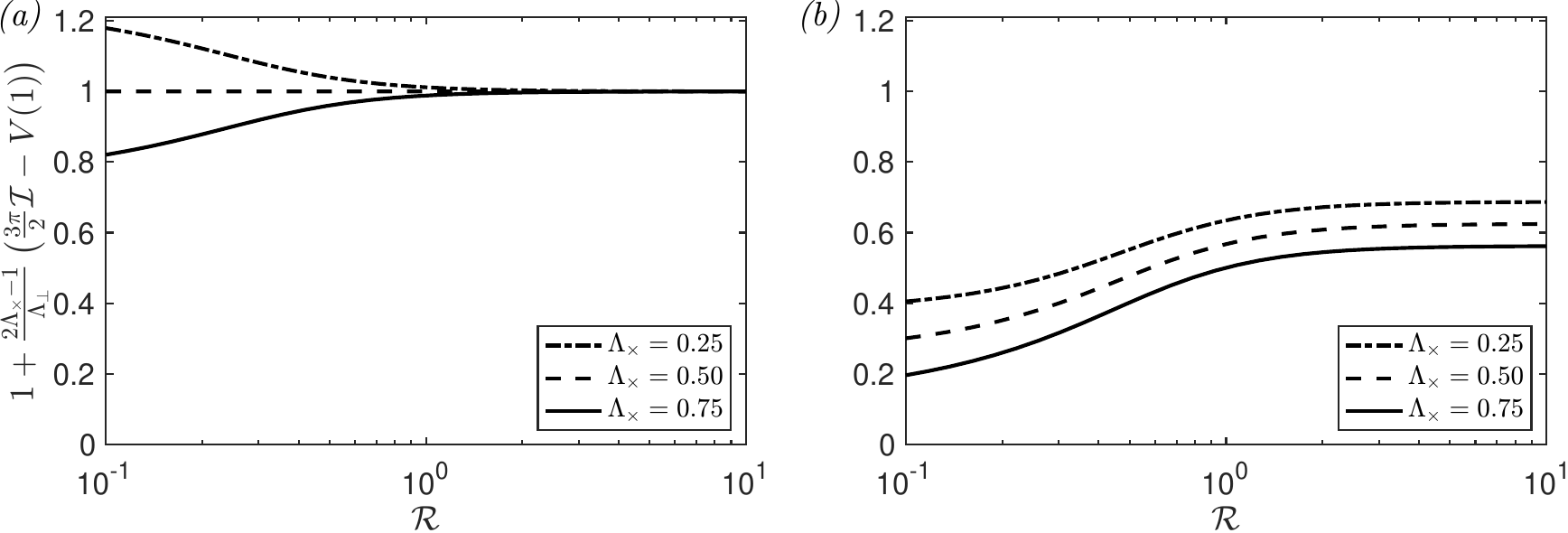}
    \caption{Dimensionless multiplier of the outward force on the parallel plates in torsion, plotted against the non-dimensional compaction length $\ndcl$ for $\Lambda_\perp=1$. \textit{(a)} The boundary-condition case of $V(1)=0$ discussed in \S\ref{sec:force-case-A}. \textit{(b)} The boundary-condition case of $\rhat\cdot\boldsymbol{\sigma}^\text{eff}(1)\cdot\rhat = 0$ discussed in \S\ref{sec:force-case-B}. For this latter case, $V(1)>0$.}
    \label{fig:force}
\end{figure}

Empirically reasonable values of the non-dimensional compaction length $\ndcl$ are $O(1)$.  Figure~\ref{fig:force}(a) shows the non-dimensional force multiplier (in square brackets in~\eqref{eq:dimensional-force-integral-general}) for $\ndcl$ in this neighbourhood and three values of $\Lambda_\times$.  Depending on whether $\Lambda_\times$ is greater than or less than 1/2, the liquid pressure (via $\mathcal{I}$) contributes negatively or positively to the excess force. Recall that for this boundary-condition case, only $\Lambda_\times < 1/2$ gives the sign of $V$ consistent with experiments. If $\Lambda_\times \lesssim 1/2$, the nondimensional force multiplier is $\sim1$, meaning the the excess force is dominated by the direct effect of dilatancy. For this boundary-condition case, we therefore approximate the normal force on the parallel plates as
\begin{equation}
    \Delta F \approx \torque D_0 \Lambda_\perp/3R.
\end{equation}
To estimate $\Delta F$, we take $D_0=3$ and $\Lambda_\perp=1$, consistent with considerations developed in the main text; we adopt values representative of laboratory experiments $R=5\times10^{-3}$~m and $\torque=10$~N~m   \citep{king2010stress}. Using these, the excess outward force exerted by the sample is about 2~kN, which is about 10\% of the force $F_c$ due to the experimental confining pressure (300~MPa).

\subsection{No radial effective stress at $r=R$}
\label{sec:force-case-B}

For the boundary condition \eqref{eq:confining-pressure-nd-bc} representing zero radial effective stress at the outer edge of the cylinder and with uniform porosity, the analytical solution for $V(r)$ is equation \eqref{eq:torsion-zeroeffstress-analytical}. The excess axial force outward on the parallel plates is given by using this solution in equation~\eqref{eq:dimensional-force-integral-general}. This case differs from that considered in \S\ref{sec:force-case-A} by the non-zero contribution of net decompaction, represented by the dimensionless $V(1)>0$. It also differs in the contribution of the liquid pressure, associated with the integral
\begin{equation}
    \mathcal{I} = \int_0^1\int_r^1 \left[ \frac{3L_0\left(\tfrac{1}{\ndcl}\right) - 2\ndcl L_1\left(\tfrac{1}{\ndcl}\right)+\tfrac{3}{\pi\ndcl}\tfrac{\Lambda_\times}{1/2-\Lambda_\times}}{3I_0\left(\tfrac{1}{\ndcl}\right) - 2\ndcl I_1\left(\tfrac{1}{\ndcl}\right)}I_1\left(\frac{r'}{\ndcl}\right) - L_1\left(\frac{r'}{\ndcl}\right)\right]\infd r'\,r\,\infd r.
\end{equation}
The liquid pressure (and hence $\mathcal{I}$) now depends on $\Lambda_\times$ due to its appearance in the boundary condition.  The dimensionless force multiplier (in square brackets in \eqref{eq:dimensional-force-integral-general}) is plotted in \ref{fig:force}(b) for three values of $\Lambda_\times$. Across the range of $\ndcl$ and $\Lambda_\times$ considered, the force multiplier ranges from about 20\% to about 60\%.  Dilatancy in the radial direction evidently reduces the outward normal force on the parallel plates.  Using the experimental and theoretical values quoted in \S\ref{sec:force-case-A}, the expected outward excess force is about 1~kN, which is about 5\% of the force due to confining pressure.


\section{Finite-time and steady models of parallel-plate torsion} \label{sec:torsion-finite-time}

By considering the system of conservation equations~\eqref{eq:conservation-equations}, we can derive a model for the evolution of the radial distribution of porosity in parallel-plate torsion.  We again assume the simplified, axisymmetric torsional flow considered in \S\ref{sec:torsion}, 
\begin{equation}
    \label{eq:torsion-flow-appendix}
    \vel^s = V(r)\rhat + \dot{\Omega}\frac{rz}{H}\phihat, \qquad
    \strr\sim\dot{\Omega}r/2H.
\end{equation}
This flow is substituted into the compaction equation~\eqref{eq:compaction}, which is then integrated subject to boundary conditions $V(R) = 0$ and $\partial P^\ell/\partial r\vert_R = 0$.  The radial component of the momentum conservation equation~\eqref{eq:stokes} is simplified with \eqref{eq:torsion-flow-appendix} and used to eliminate the pressure gradient.  We non-dimensionalise the result, along with  equation~\eqref{eq:solid-mass} and equation~\eqref{eq:non-local-viscosity} using characteristic scales 
\begin{equation}
    [r] = R,\quad [\shearvisc,\shearviscnl] = \eta_0,\quad [V] = \frac{D_0\dot\Omega R^2}{6H},\quad [t] = R/[V]
\end{equation}
to obtain 
\begin{subequations}
    \label{eq:radtor-system}
    \begin{align}
      \label{eq:radtor-vel}
      \frac{V}{\ndcl^2} &= \left(\phi/\phi_0\right)^n\left[\shearviscnl\left(\pd{}{r}\frac{1}{r}\pd{}{r}rV - (2\Lambda_\times - 1)\right) + \pd{\shearviscnl}{r}\left(\pd{V}{r} + \frac{V}{3r} - r\Lambda_\times\right)\right],  \\
      \label{eq:radtor-phi}
      \pd{\phi}{t} &+ V\pd{\phi}{r} = \left(1 -\phi\right)\frac{1}{r}\pd{}{r}rV,\\
      \label{eq:radtor-phi2}
      \shearviscnl^{-1} &= \shearvisc^{-1} + \frac{(\epsilon\ndcl)^2}{r}\pd{}{r}r\pd{}{r} (\shearviscnl^{-1}),
    \end{align}
\end{subequations}
where the dimensionless local viscosity is $\eta_\phi = \exp[-\lambda(\phi-\phi_0)]$ and $\epsilon \equiv \xi/\delta$. It is important to note that the dilation-viscosity coefficient $D_0$ does not appear in these equations; it appears only in the relationship between strain at the outer radius $\gamma$ and dimensionless time $t$,
\begin{equation}
    \label{eq:strain-time-scaling}
    t = \gamma D_0/6.
\end{equation}
This relationship indicates that a given dimensionless time (and hence amount of porosity change) is reached at smaller outer-radius strain when $D_0$ is larger. In other words, increasing $D_0$ promotes radial melt segregation.

Boundary and initial conditions imposed on the system~\eqref{eq:radtor-system} are 
\begin{subequations}
    \label{eq:radtor-bcs}
    \begin{align}
        V=0,\quad \pd{\shearviscnl}{r}=0\qquad \text{at }r=0,\\
        V=0,\quad \shearviscnl = \shearvisc \qquad \text{at }r=1,\\
        \phi(r)=\phi_0\qquad \text{at }t=0.
    \end{align}
\end{subequations}

This system \eqref{eq:radtor-system} with conditions \eqref{eq:radtor-bcs} can be solved numerically by one-dimensional finite-volume discretisation on a grid with uniform intervals in $r$ and $t$. Velocities are stored at nodes (the points that connect intervals); porosity and viscosity are stored at interval centres.  The code, developed in the framework of the Portable, Extensible Toolkit for Scientific Computation \citep[PETSc,][]{petsc-user-ref,petsc-efficient}, is available in an online repository \citep{katz2023software}.

Numerical solutions indicate that $\phi(r,t)$ tends toward a steady state in which $V(r)=0$.  This state can be determined analytically for $\epsilon=0$ and hence when $\shearviscnl=\shearvisc$. Then, \eqref{eq:radtor-vel} reduces to
\begin{equation}
    \frac{1}{\shearvisc}\pd{\shearvisc}{r} =  \frac{1 - 2\Lambda_\times }{r\Lambda_\times}.
\end{equation}
Solving and rewriting in terms of the porosity gives
\begin{equation}
    \phi(r) = \phi_0 - \frac{1-2\Lambda_\times}{\lambda\Lambda_\times}\left(\ln r + \frac{1}{2}\right),
\end{equation}
where we have determined the constant of integration using global conservation of liquid mass, $2\int_0^1r\phi\,\infd r = \phi_0$.  

\section{Cone-and-plate torsional flow}
\label{sec:cone-plate-torsion}

Cone-and-plate torsional flow is another geometry that has been used in experiments on dense granular suspensions \citep{guazzelli2018rheology}.  Although there are currently no deformation experiments on partially molten rock in this configuration, we consider the predicted radial flow for completeness. We use the flow described by \eqref{eq:torsion-velocity-ansatz} but now take the gap $H$ to be increasing linearly with radius, $H=r$. With this choice, the shear-strain rate is independent of radius.  As a consequence, the non-dimensional governing equation becomes
\begin{equation}
    \label{eq:cylindrical-cone-plate}
    \pd{}{r}\frac{1}{r}\pd{}{r}rV - \frac{V}{\ndcl^2} = \frac{\Lambda_\times - 1}{r},
\end{equation}
where we have scaled the radial velocity component $V$ with
\begin{equation}
    \label{eq:velscale-coneplate}
    [V] = \frac{D_0\dot{\Omega}R}{6}.
\end{equation}

\begin{figure}
    \centering
    \includegraphics[width=\textwidth]{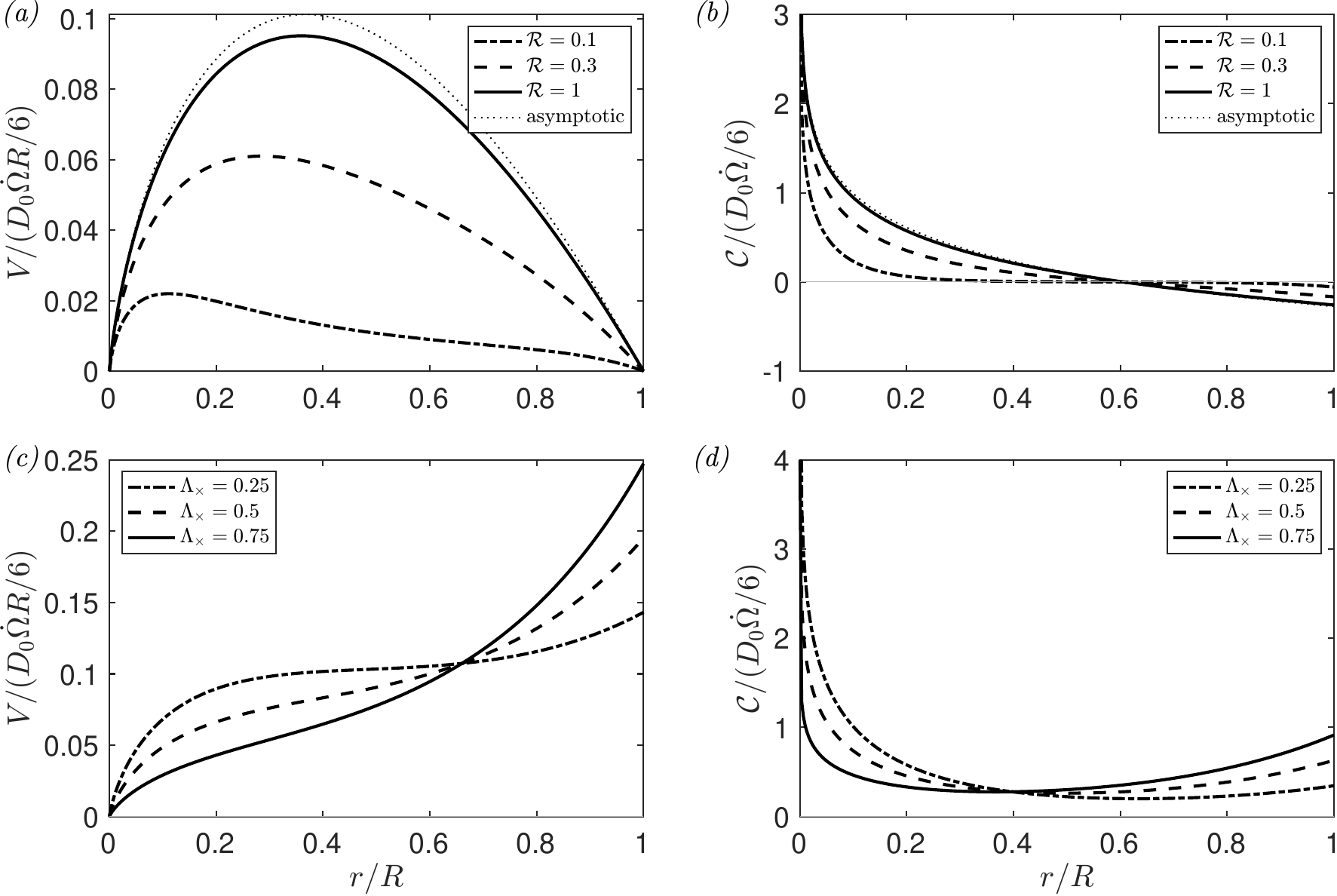}
    \caption{Cone-and-plate torsion flow. Nondimensional solutions of equation~\eqref{eq:cylindrical-cone-plate} with uniform $\eta_\phi=\eta_0$. Top panels have outer boundary condition $V(1)=0$; bottom panels have outer boundary condition as given in eqn.~\eqref{eq:confining-pressure-nd-bc}.  \textit{(a)} Analytical solutions~\eqref{eq:cone-plate-analytical-bc1} for $V$ with $\Lambda_\times=0.45$. Asymptotic solution $V(r) = (\Lambda_\times - 1)(r\ln r)/2$ for $\ndcl\to\infty$. \textit{(b)} Decompaction rate with $\Lambda_\times=0.45$. \textit{(c)} Analytical solution~\eqref{eq:cone-plate-analytical-bc2} for $V$ with $\ndcl=0.3$. \textit{(d)} Decompaction rate with $\ndcl=0.3$.}
    \label{fig:cone-plate-torsion-radial-soln}
\end{figure}

Equation \eqref{eq:cylindrical-cone-plate} with inner boundary condition $V(0)=0$ has the general solution
\begin{equation}
    V(r) = \ndcl(\Lambda_\times - 1)\left[K_1(r/\ndcl) - \ndcl/r\right] + B I_1(r/\ndcl),
\end{equation}
where $I_n(z)$ is the modified Bessel function of the first kind and $K_n(z)$ is the modified Bessel function of the second kind. The solution with outer boundary condition $V(1)=0$ is
\begin{equation}
    \label{eq:cone-plate-analytical-bc1}
    V(r) = \ndcl (\Lambda_\times - 1)\left[\frac{\ndcl - K_1(\tfrac{1}{\ndcl})}{I_1(\tfrac{1}{\ndcl})}I_1\left(\frac{r}{\ndcl}\right) +  K_1\left(\frac{r}{\ndcl}\right) -\frac{\ndcl}{r} \right].
\end{equation}
In the large-compaction-length limit of $\ndcl\gg1$, the radial velocity solution is asymptotic to $V(r) = (\Lambda_\times - 1)(r\ln r)/2$. A small-$\ndcl$ matched asymptotic solution exists but converges only for very small $\ndcl$. 

Figure~\ref{fig:cone-plate-torsion-radial-soln}(a) shows plots of solution~\eqref{eq:cone-plate-analytical-bc1} for three values of $\ndcl$ and for $\ndcl\to\infty$. We have chosen $\Lambda_\times=0.45$ for plotting purposes.  The pattern of radial flow is similar to the parallel-plate case, but with a maximum speed that is larger and shifted toward the centre of the cylinder. The associated decompaction rate (panel~(b)) increases sharply near $r=0$.  This decompaction is balanced by weak compaction near $r=1$. 

For solutions as in fig.~\ref{fig:cone-plate-torsion-radial-soln}(a) with no flow at the outer boundary, we require $\Lambda_\times < 1$ to achieve a dimensional $V>0$ consistent with experiments. This is less restrictive than the constraint~\eqref{eq:parallel-torsion-constraint} obtained from parallel-plate torsion with the same boundary conditions.

The zero-effective-stress boundary condition $\rhat\cdot\boldsymbol{\sigma}^\text{eff}\cdot\rhat = 0$ at the outer boundary, after non-dimensionalising with $[V]$ from eqn.~\eqref{eq:velscale-coneplate}, is written
\begin{equation}
    \label{eq:coneplate-confining-pressure-nd-bc}
    \frac{V}{3} + \pd{V}{r} = \Lambda_\times \quad\text{at}\quad r=1.
\end{equation}
The analytical solution in this case is 
\begin{equation}
    \label{eq:cone-plate-analytical-bc2}
    V(r) = \ndcl (\Lambda_\times - 1)\left[\frac{3K_0(\tfrac{1}{\ndcl}) + 2\ndcl K_1(\tfrac{1}{\ndcl})-2\ndcl^2 + 3\tfrac{\Lambda_\times}{\Lambda_\times-1}}{3I_0(\tfrac{1}{\ndcl}) - 2\ndcl I_1(\tfrac{1}{\ndcl})}I_1\left(\frac{r}{\ndcl}\right) +  K_1\left(\frac{r}{\ndcl}\right) -\frac{\ndcl}{r} \right].
\end{equation}
Plots of this solution with $\ndcl=0.3$ are shown in figure~\ref{fig:cone-plate-torsion-radial-soln}(c).  The radial pattern of flow and compaction are different from the zero-solid-flow boundary condition.  It is important to note that for the zero-stress boundary condition, the compaction rate need not integrate to zero over the domain because the flow (solid and liquid) at the outer boundary is non-zero.

The results in this appendix demonstrate that the radial profile of flow is sensitive to the geometry of deformation, as anticipated from previous work \citep[e.g.,][]{morris1999curvilinear}. They also show that the radial flow is sensitive to the outer boundary condition.  These results are valid at $t=0$, when the porosity, permeability, shear viscosity, and dilatancy viscosity are uniform. At later times, melt segregation causes spatial variations in porosity and hence in these coefficients.

\section{Poiseuille flow through a pipe} 
\label{sec:poiseuille}

Here we consider Poiseuille-like flow of partially molten rock along an infinite, straight pipe with circular cross-section and radius $R$. At $t=0$, the porosity is uniformly $\phi_0$. A cylindrical coordinate system $(r,\varphi,z)$ is aligned with the the axis of the pipe. The radial direction is shear-plane perpendicular and the azimuthal direction aligns with the vorticity.  An axisymmetric flow is driven by a pressure gradient $-G$ in the $z$ direction. We assume translational invariance in $z$. The solid velocity, compaction rate and deviatoric strain-rate tensor are then
\begin{equation}
    \vel^s = V(r)\rhat + W(r)\zhat, \quad \cmp = \frac{1}{r}\pd{}{r}rV, \quad \strrten = \left(\begin{array}{ccc}
        \pd{V}{r} - \frac{\cmp}{3} & 0 & \frac{1}{2}\pd{W}{r} \\
         0 & \frac{V}{r} - \frac{\cmp}{3} & 0 \\
         \frac{1}{2}\pd{W}{r} & 0 & -\frac{\cmp}{3}
    \end{array}\right).
\end{equation}
We seek a solution for $V,W$.

The $z$-component of the bulk force balance~\eqref{eq:stokes} is
\begin{equation}
    -G = \eta_0\pd{}{r}\frac{1}{r}\pd{}{r}rW.
\end{equation}
Integrating this twice subject to a symmetry condition at $r=0$ and a no-slip condition at $r=R$ gives the standard Poiseuille solution $W(r) = G\left(R^2-r^2\right)/4\eta_0$. As we did for torsion, we linearise by neglecting the contribution of dilatant flow to the second invariant of the strain rate and obtain $\strr \sim \tfrac{1}{2} \left\vert\partial{W}/\partial r\right\vert = Gr/4\eta_0$. We use this in the radial component of force-balance equation \eqref{eq:stokes} to write
\begin{equation}
    \pd{P^\ell}{r} = 3\eta_0\pd{}{r}\frac{1}{r}\pd{}{r}rV - \frac{D_0G}{4}\left(2\Lambda_\perp - \Lambda_\times\right). 
\end{equation}
Then we combine this with the compaction equation~\eqref{eq:compaction} to eliminate the liquid pressure and integrate once. Rescaling $r$ with the outer radius $R$ and $V$ with the characteristic scale
\begin{equation}
    \label{eq:radial-vel-scaling}
    [V] = \frac{G R^2 D_0}{12\eta_0} = W(0)\frac{D_0}{3},
\end{equation}
we obtain the dimensionless equation
\begin{equation}
    \label{eq:cylindrical-radial-ode}
    \pd{}{r}\frac{1}{r}\pd{}{r}rV - \frac{V}{\ndcl^2} = (2\Lambda_\perp-\Lambda_\times).
\end{equation}
Here, again, $\ndcl \equiv \sqrt{3\eta_0M_0}/R$ is the ratio of the compaction length to the outer radius. 

For a cylindrical domain with azimuthal symmetry that extends inward to $r=0$, the radial velocity must vanish on the axis, $V(0)=0$. The rigid outer wall requires that $V(1)=0$.  With these constraints, equation~\eqref{eq:cylindrical-radial-ode} admits the solution 
\begin{equation}
    \label{eq:poisseuile-zero-outer-analytical}
    V(r) = \pi\ndcl^2\left(\tfrac{\Lambda_\times}{2} -\Lambda_\perp\right)\left[ \frac{L_1(1/\ndcl)}{I_1(1/\ndcl)}I_1(r/\ndcl)-L_1(r/\ndcl)\right].
\end{equation}
Plots of this solution and its compaction rate (not shown) are identical in shape to those for torsion in figure~\ref{fig:parallel-torsion-radial-soln}, but are scaled with $(\Lambda_\times/2 - \Lambda_\perp)$ instead of $(1/2-\Lambda_\times)$. 

Results from laboratory experiments by \cite{quintanilla2019radial} that approximate Poisseuile flow indicate that $V<0$.  In light of \eqref{eq:poisseuile-zero-outer-analytical}, this requires $\Lambda_\perp > \Lambda_\times/2$. If empirical constraints from granular suspensions are relevant \citep{morris1999curvilinear, fang2002flow}, this condition is readily satisfied.  In that case, we can summarise the physics at $t=0$ as follows. Dilatant normal stress in the $r$ direction increases with radius because the shear stress increases with radius.  This pushes the solid radially inward with a stress in proportion to $\Lambda_\perp$.  The same dilatancy creates a compressive hoop stress in proportion to $\Lambda_\times$ that pushes solid radially outward. However, if condition~$\Lambda_\perp > \Lambda_\times/2$ is met, the net stress on the solid is radially inward. The solution \eqref{eq:poisseuile-zero-outer-analytical} predicts radially inward flow of the solid with decompaction at outer radii and compaction at inner radii. 

\vspace{3mm}\noindent
\textbf{Acknowledgements.} The authors thank A.~Dillman, D.~Hewitt, R.~Juanes, K.~Kamrin, D.~Kohlstedt, J.~Martin and M.~Zimmerman for helpful discussions, two anonymous reviewers for their insightful suggestions, and J.~Morris for his editorial efficiency. 

\vspace{3mm}\noindent
\textbf{Funding.} This research received funding from the European Research Council under Horizon 2020 research and innovation program grant to RFK, agreement 772255. 

\vspace{3mm}\noindent
\textbf{Declaration of interests.} The authors declare no conflict of interest.

\vspace{3mm}\noindent
\textbf{Data availability statement.} Code and data to reproduce all figures is available at \url{https://doi.org/10.5281/zenodo.10075195} \citep{katz2023software}. 

\vspace{3mm}\noindent
\textbf{Author ORCID.} R.F.~Katz \url{https://orcid.org/0000-0001-8746-5430}, J.F.~Rudge \url{https://orcid.org/0000-0002-9399-7166}, L.N.~Hansen \url{https://orcid.org/0000-0001-6212-1842}.

\vspace{3mm}\noindent
\textbf{Author contributions.} RFK conceived the study, developed the theory, analysed the theory with input from JFR, made comparison to experiments with input from LNH, and wrote the paper with input from JFR and LNH.

\bibliographystyle{jfm}
\bibliography{references}

\end{document}